\documentclass[11pt]{article}
\usepackage{amssymb, amsfonts, amsmath, amsthm}
\usepackage{geometry}
\usepackage{setspace}
\usepackage{graphicx}
\usepackage{enumitem}
\usepackage{natbib}
\usepackage{tabularx}
\usepackage{booktabs}	
\usepackage{tikz}

\newtheorem{theorem}{Theorem}
\newtheorem{assumption}{Assumption}
\newtheorem{corollary}{Corollary}
\newtheorem{lemma}{Lemma}
\newtheorem{proposition}{Proposition}
\newtheorem{remark}[theorem]{Remark}

\usepackage{hyperref}

\begin{document}

\title{Testing for Nonlinear Cointegration under Heteroskedasticity}
\author{Christoph Hanck\thanks{Address: Universit\"atsstr. 12, 45130 Essen, Germany, e-mail: \textit{christoph.hanck@vwl.uni-due.de}.}\medskip\\{\normalsize Department of Economics, University of Duisburg-Essen.}
\and Till Massing\thanks{Corresponding author. Address: Universit\"atsstr. 12, 45130 Essen, Germany, e-mail: \textit{till.massing@uni-due.de}.}\medskip\\{\normalsize Department of Economics, University of Duisburg-Essen.}}
\maketitle

\onehalfspacing

\begin{abstract}
This article discusses \citet[][Econometric Theory]{Shin1994}-type tests for nonlinear cointegration in the presence of variance breaks. We build on cointegration test approaches under heteroskedasticity \cite[][Journal of Time Series Analysis]{CavaliereTaylor2006} and nonlinearity, serial correlation, and endogeneity \cite[][Econometric Theory]{ChoiSaikkonen2010} to propose a bootstrap test and prove its consistency. A Monte Carlo study shows the approach to have satisfactory finite-sample properties in a variety of scenarios. We provide an empirical application to the environmental Kuznets curves (EKC), finding that the cointegration test provides little evidence for the EKC hypothesis. Additionally, we examine a nonlinear relation between the US money demand and the interest rate, finding that our test does not reject the null of a smooth transition cointegrating relation.
\end{abstract}

\strut


\strut


\pagebreak

\section{Introduction}\label{intro}
The concept of cointegration has proved crucial for a wide variety of empirical questions in many fields. Examples include macroeconomics, with the term structure of interest as a prominent topic. It predicts nonstationarity of the model's variables, but the existence of a stationary combination of these, thus avoiding issues of spurious regression \citep[see e.g.][for classical and recent contributions]{Phillips1986, Lin+Tu-Robuinfespurregr:20, Tu+Wang-Spurfuncregrmode:22}. There is a large literature on cointegration tests addressing a variety of possible features, such as endogeneity, serial correlation of the equilibrium errors and/or regressor innovations, heteroskedasticity and nonlinearity. For example, the environmental Kuznets curve \cite[]{Wagner2015} also discussed in our application predicts that per-capita GDP and emissions are related by an inverse U-shape as it is the poor and wealthy countries that may be expected to be, respectively, forced or capable to emit relatively little per capita. The variables may hence be modeled via a nonlinear cointegrating relation. This relation is, due to macroeconomic phenomena such as the Great Moderation, moreover plausibly affected by variance breaks. Moreover, $I(1)$ regressors in such relationships can rarely be characterized by pure random walks. Similarly, the equilibrium errors of the cointegrating relationship are often highly persistent and, moreover, correlated with the regressors' error terms. Empirical practice thus also regularly faces the need to account for serial correlation and endogeneity.

This paper presents a framework to test the null of cointegration when the cointegrating relation may be nonlinear, serially correlated, endogenous and heteroskedastic. Building on \cite{ChoiSaikkonen2010} and \cite{CavaliereTaylor2006}, both the nonlinear cointegrating relation and the variance breaks can be fairly general, the latter being allowed to occur both in the integrated regressor and in the error term.

When testing the null of \emph{no cointegration}, \cite{Engle1987} extended tests of the null of the presence of a unit root for univariate time series \cite[e.g.,][]{Dickey1979,Phillips1988} to no-cointegration tests. Alternatively, \cite{KPSS1992} test the null of stationarity against the alternative of a unit root (commonly known as KPSS test). \cite{Shin1994} extended this approach to test the null of \emph{cointegration}, as we do here. He used the ordinary least squares (OLS) residuals of a linear cointegrating regression to build the test statistic.

This theory has been developed in many directions. Some that make contributions related to our setup include \cite{Leybourne1994} and \cite{McCabe1997}, who proposed extensions of the original framework (e.g., by handling autocorrelation with a parametric adjustment).
\cite{Cavaliere2005} and \cite{CavaliereTaylor2006} incorporated variance breaks into the linear cointegration model. \cite{SaikkonenChoi2004} weakened the linearity assumption of the cointegrating regression and proposed a test for cointegrating smooth transition functions. \cite{ChoiSaikkonen2010} further extended this approach to general types of nonlinear cointegrating regressions. Both employed nonlinear least squares (NLS) estimation and leads-and-lags regression instead of OLS for estimating the potential cointegrating parameter vector.

These contributions are tremendously relevant as nonlinear cointegration has recently received increasing attention in the literature. See, for example, \citet{Wagner2015}, \citet{Stypka2017}, \citet{Hu_Phillips_Wang_2021}, \citet{Wang_Phillips_Kasparis_2021} or \citet{Lin+TuETAL-Estidoubcoin:20}. These authors address different types of nonlinearity, including polynomial and nonparametric setups. \citet{Tjstheim-Somenotenonlcoin:20} provides a partial survey. Valid testing procedures are therefore valuable.\footnote{Some contributions to the bootstrap testing literature will be reviewed in Section \ref{sec:cointtest}.}

Our main contribution therefore is to propose a test of the null of cointegration that is capable of simultaneously handling such a variety of empirically relevant features, viz.~nonlinearity, endogeneity, serial correlation and unconditional heteroskedasticity. The test thus provides a fair degree of generality and may hence be attractive to practitioners wishing to make their inference robust to a variety of data features that would otherwise make testing procedures invalid if present and not properly accounted for. Relative to \cite{CavaliereTaylor2006}, we also allow for heteroskedasticity, but additionally allow for serial correlation, endogeneity and nonlinearity. Relative to \cite{ChoiSaikkonen2010}, we also allow for nonlinearity, endogeneity, serial correlation, but additionally allow for unconditional heteroskedasticity.

We address this combination of challenges by a KPSS-type test statistic for the null of cointegration of \citet{Shin1994}. We first build on the work of \citet{ChoiSaikkonen2010} to tackle nonlinearity via suitable nonlinear least squares approaches, as well as endogeneity and serial correlation with dynamic OLS, also known as leads-and-lags regression. Second, we address unconditional heteroskedasticity via a bootstrap approach that suitable handles the fact that different patterns of heteroskedasticity imply different null distributions, making conventional tabulation of critical values impractical. Concretely, our approach to solve this problem is to draw on the fixed-regressor wild bootstrap of \citet{CavaliereTaylor2006} and establish its validity in the present, general framework. The fixed-regressor bootstrap is an attractive solution in this context as it obviates the need for the modelling of the joint dynamics of the system's variables (although we conjecture that sieve-type approaches that do so may also be feasible in this setup).

The key challenge to be met therefore is to establish the asymptotic validity of the fixed-regressor bootstrap for a KPSS-type cointegration test statistic using the residuals of an estimated nonlinear leads-and-lags cointegrating regression under the potential simultaneous presence of the above-mentioned potential complex data features. Our proofs spell out how the arguments of \citet{ChoiSaikkonen2010} (roughly, suitable linearizations to address nonlinearity and leads-and-lags to address serial correlation and endogeneity) and those of \citet{CavaliereTaylor2006} (roughly, showing that, e.g., iid standard normal wild bootstrap multipliers eliminate some nuisance parameters while replicating the variance patterns in the data so that the ``right'' null distribution is targeted by the bootstrap) can be combined to provide an asymptotically valid test under such general conditions. In particular, we establish both that the test is asymptotically level-$\alpha$ under the null and its consistency under the alternative.

The paper is organized as follows. Section \ref{sec:model} describes the nonlinear cointegrating regression model and the maintained assumptions. Section \ref{sec:cointtest} presents the cointegration tests and develops their large sample properties. In particular, we show that heteroskedasticity as well as serial correlation and endogeneity imply the presence of nuisance parameters in the asymptotic null distribution, and hence non-pivotality of the \citeauthor{Shin1994}-type (\citeyear{Shin1994}) test statistic. Hence, standard inference based on tabulated inference would be infeasible under the present, general set of assumptions. We therefore propose a dynamic regression, or leads-and-lags fixed regressor wild bootstrap to provide a feasible approach to inference given the nuisance parameters. Subsection \ref{subsec:bootstrap} shows that the bootstrap test yields asymptotically valid inference. Section \ref{sec:MCstudy} analyzes the quality of the test in a Monte Carlo study. We find that the bootstrap generally performs very well for large samples, and properly for several constellations of variance breaks with moderate differences in the rejection frequencies for sample sizes commonly considered in related work. Section \ref{sec:application} illustrates the approach with two applications, one to a panel of environmental Kuznets curves and one to the US money demand equation. We find that nonlinear cointegrating relations are not rejected for most of our series. Section \ref{sec:conclusion} concludes. Unless stated otherwise, all proofs are relegated to Appendix \ref{app:proofs}.

Some notational remarks: We denote by $\left\lfloor x\right\rfloor$ the largest integer number smaller or equal than $x\in \mathbb{R}$ and $\left\lceil x\right\rceil$ the smallest integer number larger or equal than $x$. $\Delta$ denotes the difference operator, $\textbf{1}(\cdot)$ denotes the indicator function and $\mathcal{D}_{\mathbb{R}^{m\times m}}[0,1]$ denotes the space of $m\times m$ matrices of c\`adl\`ag functions on $[0,1]$, endowed with the Skorohod topology. Weak convergence is denoted by $\stackrel{w}{\rightarrow}$, convergence in probability by $\stackrel{p}{\rightarrow}$, weak convergence in probability \cite[see][]{Gine1990} by $\stackrel{w}{\rightarrow}_p$, and almost sure convergence by $\stackrel{\mathrm{a.s.}}{\rightarrow}$. All limits are taken as $T\to\infty$, unless stated otherwise.

\section{The Model and Assumptions}\label{sec:model}
Our setup relies on a combination of the cointegration regression setup of \citet{ChoiSaikkonen2010}, allowing for nonlinearity, endogeneity and serial correlation, and the heteroskedastic setup of \citet{CavaliereTaylor2006}. This section reviews their models and assumptions and lays out how these are combined in this paper. Following \citet{ChoiSaikkonen2010}, we consider the nonlinear cointegrating regression

\begin{equation}\label{eq:cointregr}
y_t=[1,t,\ldots,t^q]'\delta+g(x_{t},\theta)+u_t,\ \ \ t=1,\ldots,T,
\end{equation}
where $y_t$ is 1-dimensional and $x_t$ is a $d$-dimensional regressor vector. We assume that $g(x_t,\theta)$ is a known smooth function of $x_t$ up to the unknown $k$-dimensional parameter vector $\theta$ and $\delta=(\delta_0,\delta_1,\ldots,\delta_q)'$. We set $\vartheta=(\delta',\theta')'$. We assume that the elements of $x_t$ are not cointegrated (see Assumption \ref{ass:spectral} below for a precise statement). This also means $g(x_t,\theta)$ is not $I(0)$, and hence that both $y_t$ and $x_t$ are $I(1)$ \citep[cf.][p.~685]{ChoiSaikkonen2010}. 

As usual, cointegration then amounts to stationarity of $u_t$. To this end, we model the error term as $u_t=\zeta_{u,t}+\mu_t$, where
\begin{equation}\label{eq:mudef}
\mu_t=\mu_{t-1}+\rho_{\mu}\zeta_{\mu,t},\ \ \ \mu_0=0.
\end{equation}

The random walk behavior of $x_t$ is specified by
\begin{equation}\label{eq:xdef}
x_t=x_{t-1}+\zeta_{x,t}.
\end{equation}
The following Assumption \ref{ass:zeta} discusses the $(d+2)$-dimensional vector process $\zeta_t:=(\zeta_{u,t},\zeta_{x,t}^{\prime},\zeta_{\mu,t})^{\prime}$.

\begin{assumption}\label{ass:zeta}
\
\begin{enumerate}[label=(\roman*)]
\item $\{\zeta_{u,t}\}$ and $\{\zeta_{\mu,t}\}$ are independent.
\item $\zeta_t:=(\zeta_{u,t},\zeta_{x,t}^{\prime},\zeta_{\mu,t})^{\prime}=\Sigma_t^{1/2}\zeta_t^*,$ where $\{\zeta_t^*\}$ is a stationary, zero-mean, unit variance process with long-run variance $\Gamma=\sum_{j=-\infty}^{\infty}E\left(\zeta_t^*(\zeta_{t-j}^*) ^{\prime}\right)$, s.th.~$\zeta_t$ is a strong-mixing sequence with mixing coefficient of size $-4r/(r-4)$, for some $r>4$ and $E||\zeta_t^*||^r<\infty$ for all $t$ and
\begin{equation}\label{eq:Omegamatrix}
\Sigma_t:=\begin{pmatrix} \sigma_{u,t}^2&\sigma_{ux,t}^{\prime}&0\\\sigma_{ux,t}&\Sigma_{x,t}&0\\0&0^{\prime}&\sigma_{\mu,t}^2\end{pmatrix}.
\end{equation}
Here, $\sigma_{u,t}^2>0$ and $\sigma_{\mu,t}^2>0$, $\sigma_{ux,t}$ is $k$-dimensional, $\Sigma_{x,t}$ ($k\times k$) is positive definite. All entries may depend on $t$. Also, $\Sigma_t$ is positive definite for any $t$.
\end{enumerate}
\end{assumption}

This means that $u_t$ has a random walk component unless $\rho_{\mu}=0$ in \eqref{eq:mudef}. Hence, the null hypothesis of cointegration is given by $H_0:\rho_{\mu}^2=0$, which is tested against the alternative $H_1:\rho_{\mu}^2>0$ of no cointegration.

Assumption \ref{ass:zeta} is similar to Assumption 1 in \cite{CavaliereTaylor2006} but additionally permits correlation between $\zeta_{u,t}$ and $\zeta_{x,t}$ to allow for endogeneity via $\sigma_{ux,t}$.\footnote{We expect that allowing for correlation between, e.g., $\zeta_{u,t}$ and $\zeta_{\mu,t}$ will not reveal additional insights. This is because a non-zero correlation between the error $u_t$ and the regressors $x_t$ in \eqref{eq:cointregr} is sufficient to capture endogeneity effects. We, therefore, abstain from considering further non-zero terms in \eqref{eq:Omegamatrix}.} Moreover, we also generalize \citet[Assumption 1]{CavaliereTaylor2006} in terms of permitting autocorrelation of the $\zeta_t$'s. This is adopted from Assumption 2 of \cite{ChoiSaikkonen2010}. 

Following \cite{Cavaliere2005} and \cite{CavaliereTaylor2006}, we allow for general forms of heteroskedastic errors via the time-varying covariance matrix $\Sigma_t$ introduced in Assumption \ref{ass:zeta}(i):

\begin{assumption}\label{ass:Omega}
The sequence $\{\Sigma_t\}_{t=1}^T$ satisfies $\Sigma_T(s):=\Sigma_{\left\lfloor Ts\right\rfloor}=\Sigma(s)$, where $\Sigma(\cdot)$ is a non-stochastic function which lies in $\mathcal{D}_{\mathbb{R}^{(d+2)\times(d+2)}}[0,1]$, with $i,j$-th element $\Sigma_{ij}(\cdot)$.
\end{assumption}

Assumption \ref{ass:Omega} allows for many possible covariance matrices of $\zeta_t$. For simple or multiple variance shifts, $\Sigma_{ij}(\cdot)$ is a piecewise constant function. For example, $\Sigma_{ij}(s):=\Sigma_{ij}^0+(\Sigma_{ij}^1-\Sigma_{ij}^0)\textbf{1}\left(s\ge\left\lfloor \tau_{ij}\right\rfloor\right)$ represents a shift from $\Sigma_{ij}^0$ to $\Sigma_{ij}^1$ at time $\left\lfloor \tau_{ij}T\right\rfloor$ $(0\le \tau_{ij}\le 1)$. Other possibilities are, e.g., affine functions ($\Sigma_{ij}(s)$ exhibits a linear trend), piecewise affine functions, or smooth transition functions. The assumption also allows for very general combinations of variance-covariance shifts. For example, the variance of $\zeta_{u,t}$ can have a shift while $\zeta_{x,t}$ is homoskedastic or heteroskedastic with a different shift function $\Sigma_{ij}(s)$. Notice that variance shifts in $\zeta_{\mu,t}$ are only relevant under the alternative $H_1$. Although we rule out stochastic volatility here, a generalization to a stochastic $\{\Sigma_t\}$, s.th.~$\{\Sigma_t\}$ is strictly exogenous w.r.t.~$\{\zeta_t^*\}$, appears possible. We refer to \cite{CavaliereTaylor2006} for details.

Furthermore, define the local long-run variance at time $t$ as $\Omega_t:=\sum_{j=-\infty}^{\infty}E\left(\zeta_t\zeta_{t-j}^{\prime}\right)$, which can be decomposed as
\begin{equation*}
\Omega_t=\begin{pmatrix} \omega_{u,t}^2&\omega_{ux,t}^{\prime}&0\\\omega_{ux,t}&\Omega_{x,t}&0\\0&0^{\prime}&\omega_{\mu,t}^2\end{pmatrix}.
\end{equation*}
This expression shall allow us to handle (co-)variance patterns that may change over time under unconditional heteroskedasticity, also known as time-varying volatility. Additionally, define
\begin{equation}\label{eq:Omegas}
\Omega(s):=\Omega_{\left\lfloor Ts\right\rfloor}.
\end{equation}
Then, the average long-run covariance matrix $\lim_{T\to\infty}\Omega_T$ is given by
\begin{equation*}
\bar{\Omega}=\int_0^1\Omega(s)\mathrm{d}s,
\end{equation*}
which can be partitioned into
\begin{equation}
\label{eq:Omegabar}
\bar{\Omega}=\begin{pmatrix} \bar{\omega}_u^2&\bar{\omega}_{ux}^{\prime}&0\\\bar{\omega}_{ux}&\bar{\Omega}_{x}&0\\0&0^{\prime}&\bar{\omega}_{\mu}^2\end{pmatrix}.
\end{equation}

Assumptions \ref{ass:zeta} and \ref{ass:Omega} imply a generalized invariance principle as stated in Lemma \ref{lem:invprinc}. The standard invariance principle as in \cite{Shin1994} would require a time-constant covariance matrix $\Sigma$. Lemma \ref{lem:invprinc} will serve as the key building block for the asymptotic distributions of the different versions of the \citet{Shin1994}-type test statistic to be presented below.

\begin{lemma}\label{lem:invprinc}
Let Assumptions \ref{ass:zeta} and \ref{ass:Omega} hold. Then,
\[
T^{-1/2}\sum_{t=1}^{\left\lfloor Ts\right\rfloor}\zeta_t\stackrel{w}{\rightarrow}B_{\Omega}(s),\ \ \ s\in[0,1],
\]
where
\begin{equation}\label{eq:BMOmega}
B_{\Omega}(s):=(B_{u,\Omega}(s),B_{x,\Omega}^{\prime}(s),B_{\mu,\Omega}(s))^{\prime}:=\int_0^s\Omega^{1/2}(r)\mathrm{d}B(r),
\end{equation}
where $B$ is a standard $(d+2)$-dimensional Brownian motion.
\end{lemma}

The next assumption \citep[cf.~Assumption 3 in][]{ChoiSaikkonen2010}, when evaluated at $\lambda=0$, ensures that the components of $x_t$ are not cointegrated. This is, as usual, necessary as \citet{Shin1994} operates in the single-equation framework dating back to at least \citet{Engle1987} where cointegration among the regressors needs to be ruled out. The typical alternative would be to work in the system-based approach pioneered by \citet{Johansen1991}.

\begin{assumption}\label{ass:spectral}
The spectral density matrix $f_{\zeta\zeta}(\lambda)$ is bounded away from zero for each $t\in\mathbb{Z}$:
\[
f_{\zeta\zeta}(\lambda)\ge \epsilon I_{d+2},\ \ \ \epsilon>0.
\]
\end{assumption}

Choosing the number $p$ in Corollary 14.3 of \cite{Davidson1994} as $2r/(r+2)$, our Assumption \ref{ass:zeta} implies the summability condition
\begin{equation}
\label{eq:summability}
\sum_{j=-\infty}^{\infty}|j|\|E(\zeta_t\zeta_{t+j}^{\prime})\|<\infty,
\end{equation}
for each $t$, which implies that the spectral density matrix is continuous \citep[again, see][for further discussion]{ChoiSaikkonen2010}.

Assumption \ref{ass:parameterandg} is the usual assumption required for deriving consistency and the asymptotic distribution of the NLS estimator.
\begin{assumption}\label{ass:parameterandg}
\
\begin{enumerate}[label=(\roman*)]
\item The parameter space $\Theta =\Theta_1\times\Theta_2$ of $\vartheta$ is a compact subset of $\mathbb{R}^k$ and the true parameter $\vartheta_0\in\Theta^0$, where $\Theta^0$ denotes the interior of $\Theta$.
\item $g(x,\theta)$ is three times continuously differentiable on $\mathbb{R}\times\Theta^*$, where $\Theta^*\supset\Theta_2$ is open.
\end{enumerate}
\end{assumption}

\section{Tests for Nonlinear Cointegration}\label{sec:cointtest}

\subsection{Roadmap}

This section develops the cointegration test that we work with in the present nonlinear setup. As is usual in the cointegration testing literature, we first need a parameter estimator based on which we obtain residuals to compute a cointegration test statistic. To this end, Section \ref{subsec:nlsreg} first provides additional assumptions required for NLS estimation of the putative cointegrating relationship's parameter vector. We then establish that the asymptotic distribution of such a standard NLS estimator depends on unknown nuisance parameters arising from endogeneity, serial correlation and heteroskedasticity under our set of general assumptions. Hence, building a test statistic on the residuals computed from such an estimator would not provide feasible inference. Here, recall that we work with the KPSS-type test statistic of \citet{Shin1994}.

Section \ref{subsec:leadsandlags} therefore introduces the dynamic, or leads-and-lags nonlinear least squares and shows that its limiting distribution is purged from the nuisance parameters arising from endogeneity and serial correlation. Our results, however, show that it is not purged from the effects of unconditional heteroskedasticity.

Section \ref{subsec:bootstrap}, the key contribution of our paper, therefore goes on to propose a fixed-regressor wild bootstrap approach. It moreover establishes the asymptotic validity of this bootstrap under the present, general set of assumptions simultaneously addressing a variety of features impacting valid inference if not accounted for properly. In particular, it shows that the bootstrap correctly replicates the asymptotic null distribution that may still be affected by nuisance parameters arising from unconditional heteroskedasticity. Concretely, we show that the bootstrap critical values provide a test with, asymptotically, the correct rejection frequency under the null as well as its consistency, i.e., a rejection probability tending to 1 under the alternative of no cointegration.

Following \cite{SaikkonenChoi2004} and \cite{ChoiSaikkonen2010} we use triangular array asymptotics in order to study the large sample behavior of the test statistic \eqref{eq:eta}, presented below. We fix the actual sample size at $T_0$ and embed the model in a sequence of models dependent on the sample size $T$, which tends to infinity. That is, we replace the regressor $x_t$ by $x_{tT}:=(T_0/T)^{1/2}x_t$ and time trends $t^j$ by $t_T^j:=(T_0/T)^{j}t^j$. This makes the regressors and regressand dependent on $T$ and we obtain the actual model for $T_0=T$. If $T_0$ is large, triangular asymptotics can be expected to give reasonable approximations to the finite sample distributions of the estimator and test statistics, see \cite{SaikkonenChoi2004}. \cite{ChoiSaikkonen2010} note that conventional asymptotic results on the NLS estimator are not available when the error term $u_t$ is allowed to be serially correlated or $x_t$ is not exogenous. See \cite{SaikkonenChoi2004} and \cite{ChoiSaikkonen2010} for a more detailed discussion of triangular asymptotics in the present context.

In particular, we embed the model \eqref{eq:cointregr} in a sequence of models
\begin{equation}\label{eq:triangarray}
y_{tT}=[1,t_T,\ldots,t_T^q]'\delta+g(x_{tT},\theta)+u_t,\ \ \ t=1,\ldots,T.
\end{equation}
As \cite{SaikkonenChoi2004} we work with an encompassing function
\[
h(t_T,x_{tT},\vartheta):=\delta_0+\delta_1t_T+\ldots+\delta_qt_T^q+g(x_{tT},\theta).
\]

In practice, we always choose $T_0=T$. We define $B_{x,\Omega}^0:= T_0^{1/2}B_{x,\Omega}$. Following the discussion of \citet[][p.~308]{SaikkonenChoi2004} we do not specify dependence on $T$ for $u_t$ \citep[as in ][p.~687 and thereafter]{ChoiSaikkonen2010}.

Again, we build on additional assumptions of \citet{ChoiSaikkonen2010} about the functions $h$ and $K$, where $K(t,x,\vartheta_0):=\left.\frac{\partial h(t,x,\vartheta)}{\partial\vartheta}\right|_{\vartheta=\vartheta_0}$, to show that, under the null, the estimators studied below are consistent and to derive their asymptotic distribution in Propositions \ref{prop:NLLasymp} and \ref{prop:LLasymp} below.

Assumption \ref{ass:gfunct} guarantees that the limit of the objective function is minimized (a.s.) at the true parameter vector $\vartheta_0$.
\begin{assumption}\label{ass:gfunct}
For some $s\in[0,1]$ and all $\vartheta\neq\vartheta_0$, $h\left(s,B_{x,\Omega}^0(s),\vartheta\right)\neq h\left(s,B_{x,\Omega}^0(s),\vartheta_0\right)\ \ \ \mathrm{(a.s.)}$.
\end{assumption}
Assumption \ref{ass:Kfunct} shall allow to establish the limiting distribution of the NLS estimator.
\begin{assumption}\label{ass:Kfunct}
$\mathcal{K}:=\int_0^1K\left(T_0s,B_{x,\Omega}^0(s),\vartheta_0\right)K\left(T_0s,B_{x,\Omega}^0(s),\vartheta_0\right)^{\prime}\mathrm{d}s>0\ \ \ \mathrm{(a.s.)}.$
\end{assumption}

\subsection{Nonlinear least squares (NLS) regression}\label{subsec:nlsreg}

We initially discuss NLS regression to estimate $\vartheta_0$. As such, the NLS estimator shall turn out not be directly useful for inferential purposes. It does provide a building block for our main suggested approach based on leads-and-lags, or dynamic regressions, to be presented in the next subsection. Let
\begin{equation}\label{eq:Qminimize}
Q(\vartheta)=\sum_{t=1}^T(y_{tT}-h(t_T,x_{tT},\vartheta))^2
\end{equation}
be the objective function to be minimized with respect to $\vartheta\in\Theta$ and $\hat{\vartheta}_T$ its minimizer.\footnote{Since $Q$ is continuous on $\Theta$ for each $(y_{1T},\ldots,y_{TT},x_{1T},\ldots,x_{TT})$ and $\Theta$ is compact by Assumption \ref{ass:parameterandg}, the NLS estimator $\hat{\vartheta}_T$ exists and is Borel measurable \cite[]{Potscher2013}.}
\begin{proposition}\label{prop:NLLasymp}
Let $K_1(x,\theta_0)=\left.\frac{\partial }{\partial x^{\prime}}\frac{\partial g(x,\theta)}{\partial\theta}\right|_{\theta=\theta_0}$ and $\kappa=\sum_{j=0}^{\infty}E(\zeta_{x,0}\zeta_{u,j})$. Then, under Assumptions \ref{ass:zeta}--\ref{ass:Kfunct} and $H_0$,
\begin{align}\label{eq:NLLasymp}
T^{1/2}\left(\hat{\vartheta}_T-\vartheta_0\right)&\stackrel{w}{\rightarrow}\mathcal{K}^{-1}\left(\int_0^1K\left(T_0s,B_{x,\Omega}^0(s),\vartheta_0\right)\mathrm{d}B_{u,\Omega}(s)+\begin{pmatrix}0\\\int_0^1K_1\left(B_{x,\Omega}^0(s),\theta_0\right)\mathrm{d}s\kappa\end{pmatrix}\right)\\
&=:\ \psi\left(B_{x,\Omega}^0,\vartheta_0,\kappa\right)\notag
\end{align}
\end{proposition}
\begin{proof}
The proof can be directly adapted from the proof of Theorem 2 in \cite{SaikkonenChoi2004} and Theorem A.1 in \cite{ChoiSaikkonen2010} combined with Lemma \ref{lem:invprinc} for $\hat\theta_T$. For $\hat\delta_T$, we use \eqref{eq:detconv} and \eqref{eq:cmtKxtT} as specified in the appendix. The zero in the second term in \eqref{eq:NLLasymp} stems from the trend regressors not correlating with $u_t$, hence not contributing to $\kappa$.
\end{proof}

Proposition \ref{prop:NLLasymp} thus generalizes Theorem A.1 in \cite{ChoiSaikkonen2010} to also allow for heteroskedasticity. However, the limiting distribution in Proposition \ref{prop:NLLasymp} is not mixed normal but involves nuisance parameters arising from both serial correlation, endogeneity and heteroskedasticity.

This subsequently translates into the limiting distribution of a test statistic using NLS residuals, which we establish next. Concretely, to test cointegration we test the stationarity of the error process $u_t$. The test is residual-based and builds on the cointegration test of \cite{Shin1994}, which, in turn, is based on the KPSS test \cite[]{KPSS1992}. Consider
\begin{equation}\label{eq:eta}
\hat{\eta}_{NLS}:=(T^2\hat{\omega}_u^2)^{-1}\sum_{t=1}^T\left(\sum_{j=1}^t\hat{u}_j\right)^2,
\end{equation}
where the $\hat{u}_t$ are the residuals for \eqref{eq:triangarray} using $\hat\vartheta_T$, the solution to \eqref{eq:Qminimize}.\footnote{Of course, these residuals, as is always the case, do depend on the sample size $T$ used for estimation, which one might make explicit via, say, $\hat{u}_{tT}$. We however omit this for notational brevity.}
That is,
\begin{equation}\label{eq:nlsresiduals}
  \hat{u}_t:=y_t-h(t,x_{t},\hat\vartheta_T)
\end{equation}
Also, $\hat{\omega}_u^2:=T^{-1}\sum_{t=1}^T\hat{u}_t^2+2T^{-1}\sum_{s=1}^lw(s,l)\sum_{t=s+1}^T\hat{u}_t\hat{u}_{t-s}$, where $w$ is a kernel which fulfills, e.g., the conditions of \cite{Andrews1991}. The lag truncation parameter $l:=l_T$ depends on the sample size such that $1/l+l/T\to0$ for $T\to \infty$. Under these conditions, $\hat{\omega}_u^2$ is a consistent estimator of $\bar{\omega}_u^2$, the $(1,1)$ element of the average long-run variance $\bar\Omega$ in \eqref{eq:Omegabar}, see also \cite{Cavaliere2005}.\footnote{\label{fn:sigmahat}The linear case without autocorrelation corresponds to the setup considered by \cite{CavaliereTaylor2006}. One could then use the standard estimator
$\hat{\sigma}_u^2:=T^{-1}\sum_{t=1}^T\hat{u}_t^2$ for the variance. In this case one can show the consistency of $\hat{\sigma}_u^2$ similarly as in \cite{CavaliereTaylor2006}.} 

Under the null, we obtain the following asymptotic behavior of $\hat{\eta}_{NLS}$.
\begin{proposition}\label{prop:etaasymp}
Under Assumptions \ref{ass:zeta}--\ref{ass:Kfunct} and $H_0$
\begin{equation}\label{eq:etaasymp}
\hat{\eta}_{NLS}\stackrel{w}{\rightarrow}\bar{\omega}_u^{-2}\int_0^1\left(B_{u,\Omega}(s)-F(s,B_{x,\Omega}^0,\vartheta_0)^{\prime}\psi(B_{x,\Omega}^0,\vartheta_0,\kappa)\right)^2\mathrm{d}s,
\end{equation}
where $F(s,B_{x,\Omega}^0,\vartheta_0):=\int_0^sK(T_0r,B_{x,\Omega}^0(r),\vartheta_0)\mathrm{d}r$ and $\psi(B_{x,\Omega}^0,\vartheta_0,\kappa)$ is defined in Proposition \ref{prop:NLLasymp}.
\end{proposition}

Note the presence of $\kappa$ in \eqref{eq:etaasymp}, so that endogeneity will affect $\hat{\eta}_{NLS}$, again illustrating the inapplicability of NLS for inferential purposes. In particular, we are not aware of how to construct a bootstrap procedure accounting for $\kappa$. We hence next proceed to establish that the dynamic regression, or leads-and-lags estimator, yields a limiting distribution only affected by heteroskedasticity. We will, in Section \ref{subsec:bootstrap}, then show this distribution to be amenable to a suitable wild bootstrap.

\subsection{Dynamic nonlinear least squares}\label{subsec:leadsandlags}

Proposition \ref{prop:NLLasymp} illustrates the well-known fact that the presence of endogeneity causes a bias through $\kappa$. Potential remedies include fully modified OLS \cite[]{Phillips1990} as suggested in \cite{Wagner2016}. We here describe the dynamic nonlinear least squares (DNLS) estimator or leads-and-lags estimator for nonlinear cointegrating regressions proposed by \cite{ChoiSaikkonen2010}. It extends the dynamic ordinary least squares (DOLS) estimator by \cite{Saikkonen1991} for linear cointegration.

First, we introduce some notation. Given our assumptions, the error term $u_t$ has the decomposition
\begin{equation}\label{eq:errordecomp}
u_t=\sum_{j=-\infty}^{\infty}\pi_j^{\prime}\zeta_{x,t-j}+e_t,
\end{equation}
where $e_t$ is a zero-mean linear projection error which is $I(0)$ under the null such that $E(e_t\zeta_{x,t-j})=0$ for all $j\in\mathbb{Z},t\in\mathbb{Z}$, and
\begin{equation}
\label{eq:pisumability}
\sum_{j=-\infty}^{\infty}(1+|j|)\|\pi_j\|<\infty.
\end{equation}
Then, the long-run variance of $e_t$ is given by $\omega_{e,t}^2 = \omega_{u,t}^2-\omega_{ux,t}^{\prime}\Omega_{x,t}\omega_{ux,t}$, where the terms on the RHS are the long-run variances in \eqref{eq:Omegas}. Similarly, the variance of $e_t$ is $\sigma_{e,t}^2 = \sigma_{u,t}^2-\sigma_{ux,t}^{\prime}\Sigma_{x,t}\sigma_{ux,t}$. Then, we may write $e_t=\sigma_{e,t}e_t^*$, where $\{e_t^*\}$ is stationary, zero-mean, unit variance, with long-run variance
\begin{equation}
\label{eq:gammae}
\gamma_e:=\sum_{j=-\infty}^{\infty}E\left(e_t^*(e_t^*)^{\prime}\right).
\end{equation}
Analogously to \eqref{eq:Omegas}, define $\omega_{e}^2(s):=\omega_{e,\left\lfloor sT\right\rfloor}^2$. Furthermore, the average long-run variance of $e_t$ is
\[
\bar\omega_{e}^2 := \int_0^{1}\omega_e^2(s)\mathrm{d}s = \bar\omega_{u}^2-\bar\omega_{ux}^{\prime}\bar\Omega_{x}\bar\omega_{ux}.
\]
We now describe the DNLS estimator. To do so, plug \eqref{eq:errordecomp} and \eqref{eq:xdef} into \eqref{eq:cointregr} to obtain
\begin{equation}\label{eq:cointregrleads}
y_t = h(t,x_t,\vartheta)+\sum_{j=-K}^K\pi_j^{\prime}\Delta x_{t-j}+e_{Kt},\ \ \ t=K+2,K+3,\ldots,
\end{equation}
with $e_{Kt}=e_t+\sum_{|j|>K}\pi_j^{\prime}\zeta_{x,t-j}$. As for the NLS regression \eqref{eq:triangarray}, we use triangular asymptotics and embed \eqref{eq:cointregrleads} into the sequences of models defined by
\[
y_{tT}=h(t_T,x_{tT},\vartheta)+V_t^{\prime}\pi+e_{Kt},\ \ \ t=K+2,\ldots,T-K,
\]
where $x_{tT}=(T_0/T)^{1/2}x_t$, $t_T=(T_0/T)t$, $V_t=\left(\Delta x_{t-K}^{\prime},\ldots,\Delta x_{t+K}^{\prime}\right)^{\prime}$ and $\pi=\left(\pi_{-K}^{\prime},\ldots,\pi_K^{\prime}\right)^{\prime}$.

Recall that $\hat\vartheta_T$ is the NLS estimator, i.e., the solution to \eqref{eq:Qminimize}. We follow \citet{ChoiSaikkonen2010} in defining the DNLS estimator as a two-step estimator using $\hat\vartheta_T$ as the first step. More precisely, the DNLS estimator is
\[
\begin{pmatrix} \hat\vartheta_T^{(1)} \\ \hat\pi_T^{(1)} \end{pmatrix} =\begin{pmatrix} \hat\vartheta_T \\ 0 \end{pmatrix} + \left(\sum_{t=K+2}^{T-K}\hat{p}_{tT}\hat{p}_{tT}^{\prime}\right)^{-1}\sum_{t=K+1}^{T-K}\hat{p}_{tT}^{\prime}\hat{u}_t,
\]
where $\hat{u}_t=y_{tT}-h(t_T,x_{tT},\hat\vartheta_T)$ and $\hat{p}_{tT}=\left(K(t_T,x_{tT},\hat\vartheta_T)^{\prime}, V_t^{\prime}\right)^{\prime}$ with $K (t_T,x_{tT},\hat\vartheta_T) = \left.\frac{\partial h(t_T,x_{tT},\vartheta)}{\partial \vartheta}\right|_{\vartheta=\hat\vartheta_T}$.

We next establish the large sample behavior of $\hat{\vartheta}_T^{(1)}$. In particular, controlling for lagged and lead differences $\Delta x_{t\pm j}$ in the DNLS estimator is the key device to remove the bias term present in Proposition \ref{prop:NLLasymp}; also note that integration is now with respect to $\mathrm{d}B_{e,\omega}(s)$ instead of $\mathrm{d}B_{u,\Omega}(s)$. Thus, the asymptotic distribution in Proposition \ref{prop:LLasymp}, while still affected by heteroskedasticity via $\Omega$, no longer is affected by endogeneity and serial correlation.
\begin{proposition}\label{prop:LLasymp}
Suppose that Assumptions \ref{ass:zeta}--\ref{ass:Kfunct} and $H_0$ hold. Then, for $K\to\infty$ and additionally assuming that $K^3/T\to\infty$ and $T^{1/2}\sum_{|j|>K}||\pi_j||\to0$,
\begin{align}\label{eq:LLasymp}
T^{1/2}\left(\hat{\vartheta}_T^{(1)}-\vartheta_0\right)&\stackrel{w}{\rightarrow}\, \mathcal{K}^{-1}\int_0^1K\left(T_0s,B_{x,\Omega}^0(s),\vartheta_0\right)\mathrm{d}B_{e,\omega}(s)\\
&=:\, \chi_{\omega}\left(B_{x,\Omega}^0,\vartheta_0\right),\label{eq:chiomegae}
\end{align}
where $B_{e,\omega}(s):=\int_0^s\omega_e(r)\mathrm{d}B_e(r)$ and $B_e$ is a standard one-dimensional Brownian motion independent of $B$ from \eqref{eq:BMOmega}.
Also, $||\hat\pi_T^{(1)}-\pi_0||=O_p(K^{1/2}/N^{1/2})$, where $\pi_0=\left(\pi_{-K0}^{\prime},\ldots,\pi_{K0}^{\prime}\right)^{\prime}$ and the $\pi_{j0}$'s denote the true parameters.
\end{proposition}
\begin{proof}
The proof can be adapted from the proof of Theorem 3 in \cite{SaikkonenChoi2004} and Theorem A.2 in \cite{ChoiSaikkonen2010} by replacing the invariance principle with the generalized invariance principle from Lemma \ref{lem:invprinc}.
\end{proof}

Instead of using the NLS residuals $\hat{u}_t$ from \eqref{eq:nlsresiduals}, it is therefore attractive to build a test statistic using DNLS residuals
\begin{equation}
\label{eq:LLresiduals}
\hat{e}_{Kt}:=y_{tT}-h(t_T,x_{tT},\hat{\vartheta}_T^{(1)})-V_t^{\prime}\hat\pi_T^{(1)},\ \ \ t=K+2,\ldots,T-K.
\end{equation}
The KPSS-type dynamic regression-based test statistic is now defined by
\begin{equation}
\label{eq:etaLL}
\hat{\eta}_{DNLS}:=(N^2\hat{\omega}_e^2)^{-1}\sum_{t=K+2}^{T-K}\left(\sum_{j=K+2}^t\hat{e}_{Kj}\right)^2,
\end{equation}
where $\hat{\omega}_e^2$ is a consistent estimator of $\bar\omega_{e}^2$ using the residuals $\{\hat{e}_{Kt}\}$ and $N:=T-2K-1$.

Theorem \ref{thm:etaLLasymp} provides the asymptotic distribution of $\hat{\eta}_{DNLS}$. It again contains a generalization of Lemma A.3 in \cite{ChoiSaikkonen2010}, additionally allowing for heteroskedasticity. It provides a core building block for the key contribution of this paper---the bootstrap procedure to be discussed in the next subsection. Concretely, it establishes a limiting distribution of the DNLS test statistic that is, while still dependent on the specific unknown shape of the time-varying heteroskedasticity via $\Omega$, purged from the influence of serial correlation and endogeneity (note that \eqref{eq:etaLLasymp} no longer depends on $\kappa$).
\begin{theorem}\label{thm:etaLLasymp}
Under Assumptions \ref{ass:zeta}--\ref{ass:Kfunct} and $H_0$
\begin{equation}\label{eq:etaLLasymp}
\hat{\eta}_{DNLS}\stackrel{w}{\rightarrow}\bar{\omega}_e^{-2}\int_0^1\left(B_{e,\omega}(s)-F(s,B_{x,\Omega}^0,\vartheta_0)^{\prime}\chi_{\omega}(B_{x,\Omega}^0,\vartheta_0)\right)^2\mathrm{d}s,
\end{equation}
where $F(s,B_{x,\Omega}^0,\vartheta_0):=\int_0^sK(T_0r,B_{x,\Omega}^0(r),\vartheta_0)\mathrm{d}r$ and $\chi_{\omega}(B_{x,\Omega}^0,\vartheta_0)$ is defined in Proposition \ref{prop:LLasymp}.
\end{theorem}

As $\Sigma(s)$ and thus $\Omega(s)$ are generally unknown, we see that the limiting distribution depends on a variance profile with nuisance parameters, which makes tabulating critical values impractical. The bootstrap, discussed in Section \ref{subsec:bootstrap}, is a natural solution, as bootstrap methods are especially beneficial in situation with nuisance parameters, see, e.g., \cite{Efron1987}. That is, the distribution affected by the variance profile can be estimated through the bootstrap. Another potential variant would be to account for the (estimated) variance profile (cf., e.g., \eqref{eq:empvp} in Section \ref{sec:application} below) so as to restore conventional asymptotic distributions as in, e.g., \citet{Cavaliere2008}.

Under the alternative, asymptotic theory becomes more involved. Since the NLS estimator $\hat\vartheta_{T}$ is not consistent anymore \cite[see][for the linear case]{Phillips1986} a limiting distribution is hard to derive. We may, however, establish the order of magnitude of $\hat{\eta}_{DNLS}$ under $H_1$. As the following subsection shows, this turns out to be sufficient to show consistency of the cointegration test.
\begin{theorem}\label{thm:aympalternative}
Under Assumptions \ref{ass:zeta}--\ref{ass:Kfunct} and $H_1$, $\hat{\eta}_{DNLS}=O_p(T/l)$.
\end{theorem}

\subsection{Bootstrap procedure}\label{subsec:bootstrap}
We adopt a bootstrap solution to provide feasible inference building on \citeauthor{CavaliereTaylor2006}'s \citeyearpar{CavaliereTaylor2006} bootstrap test for linear cointegration in the presence of variance breaks. They use the \emph{heteroskedastic fixed regressor bootstrap} by \cite{Hansen2000}. It treats the regressors as fixed, without imposing strong assumptions on the data generating process (DGP). Theorem \ref{thm:bootstrap} below shows that the fixed regressor bootstrap replicates the correct asymptotic distribution of the DNLS-based test statistic.
We extend \cite{CavaliereTaylor2006} by also allowing for nonlinearity, serial correlation and endogeneity. We do so by using the DNLS residuals \eqref{eq:LLresiduals} instead of OLS residuals as in their bootstrap. As usual, the bootstrap does not replicate the finite-sample distribution of the test statistic, see \cite{Hansen2000}. However, Section \ref{sec:MCstudy} will demonstrate that the bootstrap works reasonably well in finite samples, as also observed by \cite{CavaliereTaylor2006} in the case without serial correlation, endogeneity and nonlinearity. Popular other bootstraps, e.g., block resampling \cite[]{Lahiri1999}, are not applicable because the regressor is integrated and heteroskedastic and the error term is potentially heteroskedastic under the null hypothesis.

\citet{Chang/ParkETAL:06} and, more recently, \citet{Reichold+Jentsch-BootInfeCoinRegr:23} employ a sieve-based VAR bootstrap that is shown to be applicable in a (albeit linear) cointegrating regression setup related to ours. The sieve amounts to a VAR fitted to the residuals of the cointegrating regression and the first differences of the regressors, whose residuals are resampled in turn. While this bootstrap is not designed to handle heteroskedasticity, a variant where the VAR residuals are subjected to a wild bootstrap multiplier suggests itself. \citet{Lee2012} illustrate that a sieve bootstrap can improve the size for the KPSS test. Section \ref{subsec:sieve} uses this approach for the test of cointegration and provides a comparison to our proposal. A comprehensive analysis of the sieve bootstrap test for cointegration is left for future research.

Among the large variety of additional existing bootstrap procedures, we mention the work of \citet{Demetrescu+Hanck-RobuInfeNearRoot:16} and \citet{Rho_Shao_2019}, who design (linear) bootstrap unit root tests to accommodate heteroskedasticity. The latter exploits the dependent wild bootstrap (DWB) of \citet{Shao-DepeWildBoot:10}, which, by itself, was designed for stationary processes and hence is not directly useful for bootstrapping nonstationary or cointegrated time series. \citet{Rho_Shao_2019} however show how to modify the DWB to make it applicable to nonstationary unit root processes. We hence conjecture that it might be possible to also design bootstrap cointegration tests exploiting the DWB.\footnote{We are indebted to an anonymous referee for this suggestion.}

The heteroskedastic fixed regressor bootstrap we employ works as follows:

\begin{enumerate}
\item Run the original DNLS regression, save residuals $\hat{e}_{Kt}$ from \eqref{eq:LLresiduals} and compute the test statistic $\hat{\eta}_{DNLS}$ as given in \eqref{eq:etaLL}.
\item Construct the bootstrap sample $y_{tT}^b:=e_t^b:=\hat{e}_{Kt}z_t$, $t=1,\ldots, T$, where $\{z_t\}$ is a sequence of i.i.d.~standard normal variates.
\item Estimate $\hat{\vartheta}_T^{(1),b}$ and $\hat\pi_T^{(1),b}$ via DNLS of $y_{tT}^b$ on $h(t_T,x_{tT},\vartheta)$, save the bootstrap residuals $\hat{e}_{Kt}^b:=y_{tT}^b-h(t_T,x_{tT},\hat{\vartheta}_T^{(1),b})-V_t^{\prime}\hat\pi_T^{(1),b}$ and compute the bootstrap test statistic as
\[
\hat{\eta}_{DNLS}^b:=(N^2(\hat{\omega}_{e}^b)^2)^{-1}\sum_{t=K+2}^{T-K}\left(\sum_{j=K+2}^t\hat{e}_{Kj}^b\right)^2,
\]
where $(\hat{\omega}_{e}^b)^2$ is the long-run variance estimate using the bootstrap sample.
\item Repeat steps 2 and 3 independently $B$ times and, given that we reject for large values, compute the simulated bootstrap $p$-value $\tilde{p}_T^b:=1-\tilde{G}_T^b(\hat{\eta}_{DNLS})$, where $\tilde{G}_T^b$ is the empirical cumulative distribution function of the bootstrap test statistics $\{\hat{\eta}_{DNLS}^b\}_{b=1}^B$.
\end{enumerate}

The decision rule then is to reject the null hypothesis at level $\alpha$ if $\tilde{p}_T^b<\alpha$.

The replications, for $B$ sufficiently large, approximate $G_T^b$, the theoretical cumulative distribution function of $\hat{\eta}_{DNLS}^b$. The associated bootstrap $p$-value is defined as $p_T^b:=1-G_T^b(\hat{\eta}_{DNLS})$. Then, as $B\to\infty$, $\tilde{p}_T^b\stackrel{\mathrm{a.s.}}{\rightarrow}p_T^b$ via the law of large numbers.\label{bootstraplln}

Using the NLS residuals from \eqref{eq:nlsresiduals} in steps 1 and 2 instead would not take into account possible endogeneity. However, we also compare the NLS-based bootstrap with the DNLS-version in our simulation study. That is, we then run a NLS regression and compute $\hat{\eta}_{NLS}$ and $\hat{\eta}_{NLS}^b$ from \eqref{eq:eta} instead of the DNLS counterparts in the above algorithm.

The next theorem shows that (i) the DNLS-based bootstrap replicates the correct asymptotic null distribution. Part (ii) provides the key result to establish that the bootstrap test is consistent (cf.~Corollary \ref{cor:bootstrap}(ii) below).

\begin{theorem}\label{thm:bootstrap}
If Assumptions \ref{ass:zeta}--\ref{ass:Kfunct} hold, then
\begin{enumerate}[label=(\roman*)]
\item under $H_0$,
\[
\hat{\eta}_{DNLS}^b\stackrel{w}{\rightarrow}_p \bar{\omega}_e^{-2}\int_0^1\left(B_{e,\omega}(s)-F(s,B_{x,\Omega}^0,\vartheta_0)^{\prime}\chi_{\sigma}(B_{x,\Omega}^0,\vartheta_0)\right)^2\mathrm{d}s,
\]
where
\begin{equation}\label{eq:chisigma}
\chi_{\sigma}\left(B_{x,\Omega}^0,\vartheta_0\right):=\mathcal{K}^{-1}\int_0^1K\left(B_{x,\Omega}^0(s),\vartheta_0\right)\mathrm{d}B_{e,\sigma}(s)
\end{equation}
\item under $H_1$, $\hat{\eta}_{DNLS}^b=O_P(1)$.
\end{enumerate}
\end{theorem}

We refer to the proof of the following Corollary \ref{cor:bootstrap} for further intuition regarding the role of the process $B_{e,\sigma}(s)$ in \eqref{eq:chisigma} relative to $B_{e,\omega}(s)$ in \eqref{eq:LLasymp}. Corollary \ref{cor:bootstrap}(i) furthermore implies that the decision rule stated below the bootstrap algorithm provides an asymptotic level-$\alpha$ test. Part (ii) establishes the consistency of the test:

\begin{corollary}\label{cor:bootstrap}
Under the assumptions of Theorem \ref{thm:bootstrap},
\begin{enumerate}[label=(\roman*)]
\item under $H_0$, $p_T^b\stackrel{w}{\rightarrow}\mathcal{U}[0,1]$,
\item under $H_1$, $p_T^b\stackrel{p}{\rightarrow}0$.
\end{enumerate}
\end{corollary}

\begin{remark}\label{rem:bootstrap}
The reason we define the bootstrap data $y_{tT}^b$ as $e_t^b:=\hat{e}_{Kt}z_t$ is that the residuals $\hat{e}_{Kt}=y_{tT}-h(t_T,x_{tT},\hat{\vartheta}_T^{(1)})-V_t^{\prime}\hat\pi_T^{(1)}$ are invariant of the value of $\vartheta$ in \eqref{eq:triangarray}. Without loss of generality we can hence set $h(t_T,x_{tT},\vartheta)=0$ in the generation of the bootstrap data, see also \cite{CavaliereTaylor2006} and \cite{Georgiev2018}.
\end{remark}

\section{Monte Carlo Study}\label{sec:MCstudy}
This section provides evidence that the proposed nonlinear cointegration test works reasonably well in small samples. We study the proposed bootstrap test for linear (Section \ref{subsec:MClinear}), polynomial (Section \ref{subsec:MCpoly}), smooth transition (Section \ref{subsec:smooth}), and threshold cointegration (Section \ref{subsec:TAR}). Sections \ref{app:simulationstrend} and \ref{subsec:sieve} provide additional exploratory simulations for the trend case as well as for a potential alternative sieve bootstrap scheme.

We compare the empirical rejection rates with those of the standard test using the tabulated critical values by \cite{Shin1994} and also with a bootstrap using just the NLS-residuals.\footnote{We also experimented with the subresidual test of \cite{ChoiSaikkonen2010}. It however performed less well than the variants presented here, so that we do not present results for brevity.} Our DGP extends the design of \cite{CavaliereTaylor2006}, who generated data with a linear cointegration relation under variance breaks, by also considering nonlinear cointegration. We still start with the linear case.

\subsection{Linear regression model}\label{subsec:MClinear}
We consider the DGP
\begin{align}
y_t&=x_t+u_t, \ \ \ t=1,\ldots,T,\label{eq:linmodel1}\\
u_t&=\nu_t+\mu_t, \ \ \ u_0=0,\label{eq:linmodel2}\\
\nu_t&=\rho \nu_{t-1}+\zeta_{u,t},\ \ \ \nu_0=0,\notag\\
\mu_t&=\mu_{t-1}+\rho_{\mu}\zeta_{\mu,t}, \ \ \ \mu_0=0,\notag\\
x_t&=x_{t-1}+\zeta_{x,t},\ \ \ x_0=0,\label{eq:linmodel4}
\end{align}
where $\zeta_t:=(\zeta_{u,t},\zeta_{x,t},\zeta_{\mu,t})^{\prime}=\Sigma_t^{1/2}\zeta_t^{*}$, $\zeta_t^*\sim N(0,I_3)$, i.i.d., $|\rho|<1$ and
\[
\Sigma_t:=\begin{pmatrix} \sigma_{u,t}^2&\sigma_{ux,t}&0\\\sigma_{ux,t}&\sigma_{x,t}^2&0\\0&0^{\prime}&\sigma_{\mu,t}^2.\end{pmatrix}
\]
In particular, we initially consider the case of a simple linear cointegrating regression with a single integrated regressor, following \cite{CavaliereTaylor2006}.

We consider abrupt variance breaks of the form
\begin{align*}
\sigma_{u,t}^2&=\sigma_{u,0}^2+(\sigma_{u,1}^2-\sigma_{u,0}^2)\textbf{1}\left(t\ge\left\lfloor \tau_{u}T\right\rfloor\right)\\
\sigma_{x,t}^2&=\sigma_{x,0}^2+(\sigma_{x,1}^2-\sigma_{x,0}^2)\textbf{1}\left(t\ge\left\lfloor \tau_{x}T\right\rfloor\right)\\
\sigma_{\mu,t}&=\sigma_{\mu,0}^2+(\sigma_{\mu,1}^2-\sigma_{\mu,0}^2)\textbf{1}\left(t\ge\left\lfloor \tau_{\mu}T\right\rfloor\right).
\end{align*}
In all simulations we set $\sigma_{u,0}^2=\sigma_{x,0}^2=\sigma_{\mu,0}^2=1$.

As noted by \cite{CavaliereTaylor2006}, under the null hypothesis $\rho_{\mu}^2=0$ four cases can occur: (i) if $\tau_{u}=\tau_{x}=0$, then $y_t$ and $x_t$ are both standard $I(1)$ processes with homoskedastic increments and cointegrated; (ii) if $\tau_{u}\neq0, \tau_{x}=0$ the permanent shocks to the system are homoskedastic (i.e., $x_t$ is integrated with homoskedastic innovations) but there is a variance shift in both the transitory component of $y_t$ and in the cointegrating relation; (iii) if $\tau_{u}=0,\tau_{x}\neq0$, the permanent shocks to the system are heteroskedastic with changes to both $x_t$ and $y_t$ being heteroskedastic, but there are no variance shifts in the cointegrating relation; (iv) if $\tau_{u}\neq0, \tau_{x}\neq0$, the permanent shocks to the system are heteroskedastic, changes to both $x_t$ and $y_t$ are heteroskedastic and there is a variance shift both in the transitory component of $y_t$ and in the cointegrating relation. If $H_0$ holds, variance shifts in $\zeta_{\mu}$ have no influence. Under the alternative we also allow for variance breaks in $\zeta_{\mu}$ which lead to variance breaks in $u_t$ which are similar to cases (ii) and (iv).

Moreover, we consider covariance breaks of the form
\[
\sigma_{ux,t}=\sigma_{ux,0}+(\sigma_{ux,1}-\sigma_{ux,0})\textbf{1}\left(t\ge\left\lfloor \tau_{ux}T\right\rfloor\right).
\]

In our simulations we only consider the case where all shifts occur at the same time, i.e., $\tau:=\tau_{u}=\tau_{x}=\tau_{\mu}=\tau_{ux}$. For the results on other possible scenarios see the simulation study of \cite{CavaliereTaylor2006} who did not observe qualitative differences for their bootstrap test.

We investigate the following parameter constellations. Let the sample size be $T\in\{100,300\}$. We take $\rho_{\mu}^2\in\{0,0.001,0.01,0.1\}$. $\rho_{\mu}^2=0$ is to estimate size, the other constellations imply a power analysis. We consider variance breaks at $\tau\in\{0,0.1,0.5,0.9\}$. Here, $\tau=0$ corresponds to the case of no variance breaks the remaining values generate early, middle, and late variance breaks. We set the magnitude of the variance breaks as $\sigma_{1}^2=\sigma_{u,1}^2=\sigma_{x,1}^2=\sigma_{\mu,1}^2\in\{1/16,16\}$, like in \cite{CavaliereTaylor2006}. The parameter for the covariance $\sigma_{ux,t}$ are chosen in such a way that the correlation between $\zeta_{u,t}$ and $\zeta_{x,t}$ is fixed over time at $\lambda\in\{0,0.5\}$, i.e., without or with endogeneity. This implies that breaks in the variance and covariance occur jointly. The $AR(1)$ parameter of $u_t$ is $\rho\in\{0,0.5,0.8\}$. Empirical rejection rates are based on 1,000 replications and the number of bootstrap replications is $B=500$. As in \cite{ChoiSaikkonen2010} we take $K\in\{1,2,3\}$ as the leads-and-lags parameter. However, we only report the case $K=1$ for brevity as the other choices yielded qualitatively similar results. Finally, the nominal significance level is $\alpha=0.05$.

We perform the test by estimating $\vartheta$ in the linear regression $y_t$ onto $h(t,x_t,\vartheta)\equiv \theta x_t$ and using the residuals to compute $\hat{\eta}_{NLS}$ and $\hat{\eta}_{DNLS}$.\footnote{While we formulate the theory for nonlinear cointegrating regressions we for simplicity use the OLS estimator whenever possible to speed up the computations.} We use the estimator $\hat{\sigma}_u^2$ or $\hat{\sigma}_e^2$, resp., given in footnote \ref{fn:sigmahat} for $\rho=0$ and, for $\rho\neq0$, a non-parametric autocorrelation-robust estimator for the long-run variance with a Bartlett kernel and a spectral window of $\left\lfloor 4(T/100)^{0.25}\right\rfloor $ as suggested in \cite{KPSS1992}. Table \ref{tab:linear} reports empirical rejection rates (as percentages) for the different parameter constellations. Panel (a) shows the rates for the bootstrap approach using NLS and panel (b) for the bootstrap test using DNLS. For comparison, Table \ref{tab:linearShin} shows rejection rates test based on the critical value 1.199 tabulated by \cite{Shin1994} for a single regressor without trend. First, the bootstrap tests generally yields very good empirical sizes and powers. Both time (early or late) and direction (increase or decrease) of a variance break do not have a systematic impact on the rejection frequencies. For example, early downward variance breaks yield lower empirical power than early upward variance breaks, and vice versa for late variance breaks. This effect reduces with increasing $\rho_{\mu}^2$. Size distortions increase in the degree of autocorrelation. This is as expected, as size distortions are a fairly common feature when performing (cointegration) inference in the presence of strong autocorrelation or endogeneity, see, e.g., \cite{Kiefer2005} for general HAC results and \cite{ChoiSaikkonen2010} and \cite{Stypka2017} for results specific to the nonlinear cointegration literature.

For Table \ref{tab:linearShin}, we observe that, as expected from Proposition \ref{prop:etaasymp}, variance breaks affect the Shin-test. Specifically, it is oversized/undersized depending on whether there are downward/upward breaks. Its empirical power is generally lower than for the bootstrap test.

\begin{table}[tbp]
 \begin{scriptsize}
		\caption{Empirical rejection frequencies for testing the null of cointegration in the \emph{linear} regression model for various parameter constellations. All rejection rates are given as percentages. The nominal size is 5\%. Panel (a) is for the bootstrap test using NLS and panel (b) is for the bootstrap test using DNLS.}
  \hspace{-0.5cm}
    \begin{tabular}{crrrrrrrrrrrrrrrrr}
    	    \toprule
		& &       & \multicolumn{1}{r}{$T$} & \multicolumn{7}{c}{100}       & \multicolumn{7}{c}{300}    \\
		\cmidrule(l{.75em}){5-11}\cmidrule(l{.75em}){12-18}
     &     &       & \multicolumn{1}{r}{$\tau$} & 0     & \multicolumn{2}{c}{0.1}          & \multicolumn{2}{c}{0.5}   &        \multicolumn{2}{c}{0.9}    & 0     & \multicolumn{2}{c}{0.1}   &  \multicolumn{2}{c}{0.5}  &   \multicolumn{2}{c}{0.9}     \\
					\cmidrule(l{.75em}){6-7}\cmidrule(l{.75em}){8-9}\cmidrule(l{.75em}){10-11}\cmidrule(l{.75em}){13-14}\cmidrule(l{.75em}){15-16}\cmidrule(l{.75em}){17-18}
     &     &       & \multicolumn{1}{r}{$\sigma_1^2$} &       & 1/16 & 16    & 1/16 & 16    & 1/16 & 16    &       & 1/16 & 16    & 1/16 & 16    & 1/16 & 16 \\
  &  \multicolumn{1}{r}{$\rho_{\mu}^2$} & \multicolumn{1}{r}{$\rho$} & \multicolumn{1}{r}{$\lambda$} &       &       &       &       &       &       &       &       &       &       &       &       &       &  \\ \midrule
      (a) & 0     & 0     & 0     & 5.3   & 6.7   & 5.1   & 5.3   & 4.8   & 5.2   & 4.7   & 4.8   & 5.6   & 5.1   & 5.1   & 5     & 4.9   & 5.4 \\
          &       &       & 0.5   & 4.7   & 6.4   & 4.4   & 4.7   & 4.8   & 4.5   & 5.7   & 4.1   & 5.1   & 4.3   & 5.1   & 5.2   & 4.4   & 5.6 \\
          &       & 0.5   & 0     & 8.3   & 6.4   & 7.3   & 7.2   & 5.6   & 7.7   & 5.7   & 7.1   & 6.4   & 6.9   & 7.2   & 6.4   & 7.1   & 6.7 \\
          &       &       & 0.5   & 6.9   & 6.4   & 6.6   & 6.6   & 4.9   & 6.4   & 6.2   & 6.2   & 6.1   & 6     & 6.5   & 6.8   & 6.1   & 6.8 \\
          &       & 0.8   & 0     & 14    & 11.7  & 13.2  & 13    & 10.2  & 13.6  & 11.8  & 10.7  & 9.1   & 11    & 10.1  & 10.9  & 10.5  & 10.6 \\\vspace{0.15cm}
          &       &       & 0.5   & 12.7  & 12.8  & 12.1  & 13.2  & 9.6   & 12.5  & 11.9  & 10.4  & 10    & 9.6   & 9.9   & 10    & 9.7   & 11.1 \\
          & 0.001 & 0     & 0     & 15.6  & 17    & 13.6  & 13.9  & 11    & 15.2  & 11.9  & 48.4  & 38.5  & 45.8  & 41.4  & 43.1  & 50.7  & 39.4 \\
          &       &       & 0.5   & 15.4  & 17.2  & 13.9  & 13.9  & 11.1  & 14.8  & 12.7  & 49.5  & 37.8  & 46.3  & 42.5  & 43.2  & 50.5  & 38.7 \\
          &       & 0.5   & 0     & 10.5  & 11.4  & 9.6   & 10.4  & 6.9   & 10.2  & 8.4   & 25.7  & 24.1  & 23.3  & 21.7  & 20.3  & 25.7  & 20.7 \\
          &       &       & 0.5   & 10.1  & 11.5  & 9     & 9.9   & 6.4   & 9.5   & 8.8   & 25.2  & 23.2  & 22.2  & 22.4  & 20.1  & 25.4  & 21 \\
          &       & 0.8   & 0     & 14    & 12.8  & 14.1  & 14.2  & 10.9  & 14.4  & 12.4  & 16.8  & 15.8  & 14.8  & 15.2  & 13.4  & 16    & 14.5 \\
          &       &       & 0.5   & 13.5  & 14.1  & 12.6  & 13.3  & 10.3  & 13.7  & 12.2  & 15.1  & 16.2  & 14.6  & 15.3  & 13.7  & 15.3  & 14.6 \\
          & 0.01  & 0     & 0     & 50.1  & 43.3  & 47.2  & 42.3  & 42.4  & 51    & 40.1  & 87.1  & 77    & 85.2  & 79.9  & 85.6  & 87.1  & 79.9 \\
          &       &       & 0.5   & 50.9  & 43.9  & 47.3  & 43.1  & 43.8  & 51.1  & 40    & 87    & 76.5  & 85.5  & 80.2  & 85.8  & 87.3  & 80.7 \\
          &       & 0.5   & 0     & 24.5  & 25.2  & 21.9  & 21.6  & 16.9  & 24.8  & 19.5  & 54.8  & 48.9  & 52    & 49    & 49.7  & 55.3  & 47.5 \\
          &       &       & 0.5   & 23.9  & 26    & 21.5  & 21.5  & 17.2  & 23.7  & 19.7  & 55.4  & 48.5  & 51.5  & 49    & 50.4  & 55.8  & 49.4 \\
          &       & 0.8   & 0     & 18.6  & 20.3  & 18.2  & 20.1  & 13.8  & 19.7  & 15.6  & 36.1  & 34.2  & 33.3  & 32.7  & 30.7  & 38    & 31.6 \\
          &       &       & 0.5   & 18.3  & 21.7  & 17.1  & 19.3  & 12.6  & 18.6  & 16    & 37.2  & 34.7  & 33.4  & 32.4  & 30.9  & 36.1  & 32.6 \\
          & 0.1   & 0     & 0     & 84.9  & 77    & 82.3  & 77.3  & 82.2  & 84.6  & 77.3  & 98.5  & 96.3  & 98.4  & 96.1  & 98.9  & 98.7  & 97.5 \\
          &       &       & 0.5   & 84.3  & 78.4  & 82.3  & 78.1  & 82    & 85.4  & 77.2  & 98.6  & 96    & 98.5  & 96    & 98.5  & 98.6  & 97.5 \\
          &       & 0.5   & 0     & 45.3  & 43.8  & 41    & 40.8  & 36.9  & 45.1  & 39.6  & 68.3  & 65.4  & 65.6  & 65.8  & 63    & 69.1  & 62.8 \\
          &       &       & 0.5   & 45    & 45.7  & 41.1  & 41.1  & 37.4  & 45.5  & 40.8  & 68.3  & 64.8  & 65.8  & 65    & 63.8  & 68.2  & 63.2 \\
          &       & 0.8   & 0     & 36.5  & 37    & 32    & 34    & 28.2  & 36.9  & 31.9  & 62.4  & 57.9  & 58.7  & 56.5  & 56.2  & 63.5  & 54.8 \\
          &       &       & 0.5   & 35.7  & 39    & 32.4  & 33.2  & 28.1  & 36.2  & 32.1  & 62.7  & 58.4  & 59.9  & 57.6  & 55.8  & 62.6  & 56.3 \\
          &       &       &       &       &       &       &       &       &       &       &       &       &       &       &       &       &  \\
      (b) & 0     & 0     & 0     & 4.7   & 7.8   & 4.9   & 5.7   & 4.6   & 4.7   & 4.2   & 5.5   & 6.5   & 4.6   & 5.2   & 5.2   & 4.2   & 4.5 \\
          &       &       & 0.5   & 4     & 7.2   & 4.1   & 5.6   & 5.4   & 5.2   & 4.6   & 4.2   & 7.3   & 5     & 5.6   & 5     & 5.2   & 5.6 \\
          &       & 0.5   & 0     & 7.6   & 9.5   & 7.8   & 7.1   & 6.9   & 7.8   & 6.1   & 6.9   & 7.9   & 6.7   & 7.3   & 7.2   & 6.6   & 6.7 \\
          &       &       & 0.5   & 8.9   & 9.9   & 6.1   & 9.3   & 5.7   & 8.8   & 7.6   & 6.2   & 6.9   & 7     & 7.9   & 6.2   & 7.2   & 6.1 \\
          &       & 0.8   & 0     & 16.2  & 18.7  & 15.7  & 17    & 11.3  & 18.2  & 14.9  & 10.7  & 7.8   & 11.2  & 10.7  & 9.4   & 10.1  & 10 \\\vspace{0.15cm}
          &       &       & 0.5   & 16    & 19.6  & 17.5  & 15.6  & 13.9  & 15.9  & 15.6  & 9.8   & 11    & 11.2  & 13.2  & 11    & 9.9   & 8.9 \\
          & 0.001 & 0     & 0     & 16    & 22.7  & 12.6  & 15.3  & 9.7   & 14.2  & 10.5  & 47.8  & 42.5  & 47.3  & 41.1  & 44.2  & 50.4  & 41.1 \\
          &       &       & 0.5   & 17.3  & 22.9  & 15.9  & 16.5  & 12.1  & 19.7  & 15.3  & 55.4  & 45.3  & 52.1  & 48    & 46.2  & 57    & 44 \\
          &       & 0.5   & 0     & 11.7  & 16.4  & 9.8   & 11.6  & 8.2   & 9.4   & 9.3   & 25.6  & 26.4  & 24.3  & 22.3  & 20.8  & 24.9  & 24.5 \\
          &       &       & 0.5   & 12.9  & 16.2  & 11.7  & 13.2  & 8.4   & 13.2  & 10.7  & 28.5  & 28.7  & 28.9  & 26.9  & 21.7  & 30.1  & 23.5 \\
          &       & 0.8   & 0     & 17    & 20.2  & 15.4  & 16.8  & 13.7  & 17.5  & 16.3  & 15.9  & 15.5  & 14.4  & 16.3  & 15.2  & 17.7  & 16.1 \\
          &       &       & 0.5   & 17.7  & 22.9  & 15.2  & 19    & 12    & 17    & 15.9  & 18.8  & 17.3  & 14.6  & 16.1  & 13.9  & 17.8  & 17.4 \\
          & 0.01  & 0     & 0     & 50.9  & 52    & 45.7  & 42.7  & 43.6  & 50.7  & 39.2  & 86    & 79.4  & 85.7  & 78.8  & 84.3  & 88.1  & 79.8 \\
          &       &       & 0.5   & 56.4  & 57.3  & 55.1  & 48.9  & 47.4  & 58.1  & 42.1  & 90.7  & 83.3  & 89.3  & 83.7  & 88.5  & 88.5  & 84.5 \\
          &       & 0.5   & 0     & 26.4  & 33    & 23.1  & 24.1  & 20.5  & 27.1  & 23.3  & 54    & 50.7  & 52.1  & 48.4  & 49.9  & 55.2  & 50.6 \\
          &       &       & 0.5   & 28.5  & 37.4  & 27.6  & 26.6  & 21.6  & 30.5  & 24.2  & 58    & 54.6  & 55.2  & 52.9  & 52.7  & 56.4  & 51.1 \\
          &       & 0.8   & 0     & 22    & 29.4  & 22.7  & 21    & 18.3  & 21.2  & 19    & 36    & 33.7  & 32.7  & 32.1  & 32.6  & 33.4  & 33 \\
          &       &       & 0.5   & 21.1  & 29.6  & 21.7  & 25    & 18.9  & 22.4  & 19.7  & 38.9  & 36.9  & 35.2  & 37.8  & 34.9  & 39.6  & 33.2 \\
          & 0.1   & 0     & 0     & 82.2  & 83.5  & 80.8  & 75.8  & 79.3  & 85.3  & 75.8  & 98.8  & 97.2  & 97.8  & 95.6  & 98    & 98.8  & 97.6 \\
          &       &       & 0.5   & 86.5  & 87.3  & 82.6  & 80.4  & 82.8  & 86.8  & 80.4  & 98.9  & 97    & 98.5  & 97.4  & 98.3  & 99.2  & 98.5 \\
          &       & 0.5   & 0     & 51.5  & 54.6  & 43.2  & 44.1  & 39.7  & 51.6  & 45.3  & 67.8  & 68.3  & 64.8  & 63.3  & 63    & 69.1  & 64.9 \\
          &       &       & 0.5   & 52.4  & 56.1  & 45.7  & 46.1  & 44.2  & 49.4  & 49.4  & 66.9  & 67.8  & 68.2  & 65.6  & 61.6  & 69.8  & 62.9 \\
          &       & 0.8   & 0     & 38.9  & 48    & 36.2  & 38.5  & 32.2  & 41    & 36.7  & 61.4  & 57.3  & 61.3  & 57.8  & 59.9  & 62.5  & 55.9 \\
          &       &       & 0.5   & 40.4  & 49.8  & 37.8  & 39    & 37    & 43.9  & 35.3  & 62.4  & 56.9  & 57.4  & 58.9  & 55.9  & 61.8  & 56.5 \\
        \bottomrule
		\end{tabular}%
  \label{tab:linear}%
	  \end{scriptsize}
\end{table}%

\begin{table}[tbp]
 \begin{scriptsize}
		\caption{Empirical rejection frequencies for testing the null of cointegration in the \emph{linear} regression model for various parameter constellations using the \cite{Shin1994} test with the critical value 1.199.}
  \hspace{-0.5cm}
    \begin{tabular}{crrrrrrrrrrrrrrrrr}
    	    \toprule
		& &       & \multicolumn{1}{r}{$T$} & \multicolumn{7}{c}{100}       & \multicolumn{7}{c}{300}    \\
		\cmidrule(l{.75em}){5-11}\cmidrule(l{.75em}){12-18}
     &     &       & \multicolumn{1}{r}{$\tau$} & 0     & \multicolumn{2}{c}{0.1}          & \multicolumn{2}{c}{0.5}   &        \multicolumn{2}{c}{0.9}    & 0     & \multicolumn{2}{c}{0.1}   &  \multicolumn{2}{c}{0.5}  &   \multicolumn{2}{c}{0.9}     \\
					\cmidrule(l{.75em}){6-7}\cmidrule(l{.75em}){8-9}\cmidrule(l{.75em}){10-11}\cmidrule(l{.75em}){13-14}\cmidrule(l{.75em}){15-16}\cmidrule(l{.75em}){17-18}
     &     &       & \multicolumn{1}{r}{$\sigma_1^2$} &       & 1/16 & 16    & 1/16 & 16    & 1/16 & 16    &       & 1/16 & 16    & 1/16 & 16    & 1/16 & 16 \\
  &  \multicolumn{1}{r}{$\rho_{\mu}^2$} & \multicolumn{1}{r}{$\rho$} & \multicolumn{1}{r}{$\lambda$} &       &       &       &       &       &       &       &       &       &       &       &       &       &  \\ \midrule
					& 0     & 0     & 0     & 5     & 12.3  & 4.2   & 9.5   & 1.7   & 6.1   & 4.2   & 5     & 12.1  & 4     & 9.9   & 1.3   & 6.2   & 3.9 \\
          &       &       & 0.5   & 3.7   & 10.7  & 3.3   & 9.2   & 1.5   & 4.6   & 4.8   & 4     & 10.9  & 3.5   & 8     & 1.3   & 4.4   & 4.5 \\
          &       & 0.5   & 0     & 10    & 12.4  & 8.7   & 15.1  & 4.4   & 11    & 9.6   & 7.6   & 13.1  & 6.5   & 12.9  & 2.6   & 9.2   & 6.7 \\
          &       &       & 0.5   & 8.1   & 11.7  & 7.2   & 13.4  & 4.2   & 8.6   & 10.4  & 6.6   & 11.5  & 5.6   & 10.7  & 2.7   & 6.9   & 7.4 \\
          &       & 0.8   & 0     & 16.7  & 17.3  & 15.1  & 22.5  & 8.4   & 17.3  & 15.8  & 11.9  & 16.7  & 10.2  & 17.4  & 4.5   & 12.4  & 10.4 \\
          &       &       & 0.5   & 14.6  & 17.1  & 13.2  & 20    & 7.7   & 15.2  & 14.9  & 9.6   & 15.5  & 8.4   & 16.5  & 4.4   & 10.6  & 11.4 \\
          & 0.001 & 0     & 0     & 15.5  & 22.2  & 12.4  & 19.5  & 4.4   & 16.6  & 9.3   & 45.6  & 44.8  & 40.8  & 47    & 27.7  & 49    & 34.6 \\
          &       &       & 0.5   & 14.7  & 20.9  & 11.6  & 19.4  & 4.3   & 15.7  & 10.5  & 44.7  & 45    & 40.5  & 46.3  & 28    & 48.2  & 35.1 \\
          &       & 0.5   & 0     & 13.8  & 16.9  & 11.4  & 18.7  & 5.4   & 14.4  & 11.6  & 25.4  & 28.9  & 21.6  & 29    & 11.1  & 27.1  & 20.1 \\
          &       &       & 0.5   & 12.2  & 15.7  & 10.2  & 17.3  & 5.1   & 12.8  & 12.9  & 23.9  & 29.2  & 20.5  & 27.7  & 11.5  & 25.8  & 20.6 \\
          &       & 0.8   & 0     & 17.9  & 18.4  & 15.6  & 22.4  & 8.8   & 19    & 15.9  & 17    & 22.1  & 13.7  & 21.7  & 7.2   & 17.9  & 14.1 \\
          &       &       & 0.5   & 15.6  & 18.3  & 13.2  & 20.9  & 7.7   & 15.4  & 15.9  & 15.2  & 21.8  & 12.8  & 21    & 6.2   & 16.4  & 15.4 \\
          & 0.01  & 0     & 0     & 46.8  & 45    & 40.3  & 46.1  & 27.9  & 47.7  & 35.8  & 85.2  & 79.4  & 80.6  & 82.7  & 74.5  & 86    & 75.9 \\
          &       &       & 0.5   & 45.8  & 44.6  & 41.6  & 46.3  & 27.9  & 47.7  & 35.8  & 84.7  & 80    & 81.1  & 81.9  & 74    & 85.5  & 76.2 \\
          &       & 0.5   & 0     & 27    & 28.3  & 23    & 29.1  & 13.7  & 27.6  & 24.7  & 49.6  & 48    & 44.4  & 49.9  & 36.8  & 50.9  & 45.4 \\
          &       &       & 0.5   & 26.1  & 28    & 23.1  & 28.3  & 14.1  & 26.9  & 24    & 50.1  & 47.8  & 44.4  & 50.1  & 36.7  & 50.8  & 45.9 \\
          &       & 0.8   & 0     & 22.1  & 23.8  & 20    & 27.7  & 11.5  & 22.6  & 20.6  & 34.7  & 35.8  & 29.3  & 36    & 19.8  & 35.6  & 30.7 \\
          &       &       & 0.5   & 20.9  & 23.6  & 18.7  & 26.7  & 11.1  & 22.7  & 21.2  & 35.1  & 35.2  & 29.6  & 36.4  & 20    & 34.9  & 30.8 \\
          & 0.1   & 0     & 0     & 79.6  & 76.5  & 75.8  & 79.3  & 68.2  & 80.6  & 71.2  & 97.8  & 96.8  & 97.2  & 97.4  & 95.6  & 98.2  & 95.6 \\
          &       &       & 0.5   & 79.3  & 75.5  & 75.7  & 79.3  & 67.6  & 80.9  & 72.2  & 98.1  & 96.7  & 96.9  & 97    & 95.7  & 98.2  & 95.8 \\
          &       & 0.5   & 0     & 42.6  & 40.9  & 38.3  & 44.3  & 32.2  & 44    & 43.3  & 59.7  & 57.2  & 55.1  & 60.2  & 50.5  & 59.4  & 58.5 \\
          &       &       & 0.5   & 42.9  & 40.6  & 38.1  & 43.4  & 31.9  & 43.9  & 44.3  & 59.8  & 56.3  & 54.2  & 61.4  & 50.8  & 59.7  & 57.9 \\
          &       & 0.8   & 0     & 36.9  & 35.4  & 32.1  & 38    & 24.4  & 36.3  & 36.2  & 53.5  & 51.9  & 49.2  & 54.1  & 43    & 54    & 51.9 \\
          &       &       & 0.5   & 35.9  & 33.8  & 31.7  & 37.7  & 25.2  & 36.2  & 36.1  & 54.8  & 51.4  & 48    & 55.3  & 43.4  & 55.6  & 51.6 \\
        \bottomrule
		\end{tabular}%
  \label{tab:linearShin}%
	  	  \end{scriptsize}
\end{table}%

Since there are some size distortions for the small samples especially in cases of both endogeneity and autocorrelation we now discuss empirical sizes for growing $T$. Table \ref{tab:linearlong} reports empirical sizes for $T\in\{500,1000,2000,3000,5000\}$. In all scenarios, the empirical size converges to the nominal size of 5\%, which illustrates that the bootstrap asymptotically performs as desired, in line with our theoretical results.

\begin{table}[tbp]
	\centering
 \begin{scriptsize}
		\caption{For $\rho_{\mu}^2=0$, empirical sizes for testing the null of cointegration in the \emph{linear} regression model for various parameter constellations. All rejection rates are given as percentages. The nominal size is 5\%.}
      \begin{tabular}{rrrrrrrrrr}
    	    \toprule
          &       & \multicolumn{1}{r}{$\tau$} & 0     & \multicolumn{2}{c}{0.1}          & \multicolumn{2}{c}{0.5}   &        \multicolumn{2}{c}{0.9}    \\
					\cmidrule(l{.75em}){5-6}\cmidrule(l{.75em}){7-8}\cmidrule(l{.75em}){9-10}
          &       & \multicolumn{1}{r}{$\sigma_1^2$} &       & 1/16 & 16    & 1/16 & 16    & 1/16 & 16     \\
 $T$  & \multicolumn{1}{r}{$\rho$} & \multicolumn{1}{r}{$\lambda$} &       &       &       &       &       &       &       \\ \midrule
    500   & 0     & 0     & 6.5   & 4.3   & 4.2   & 3.4   & 4.9   & 5.4   & 5.7 \\
          &       & 0.5   & 4.7   & 6     & 5     & 4.2   & 4.2   & 4.6   & 3.9 \\
          & 0.5   & 0     & 8     & 6.6   & 6     & 5.5   & 7     & 6.4   & 6.9 \\
          &       & 0.5   & 6.4   & 6.8   & 6.8   & 6.2   & 6     & 6.2   & 6.5 \\
          & 0.8   & 0     & 10    & 8.2   & 11.8  & 8.5   & 10.2  & 9.3   & 10.4 \\
          &       & 0.5   & 10.3  & 8.2   & 9.4   & 8.9   & 7.9   & 11    & 11.3 \\
    1000  & 0     & 0     & 4.5   & 4.8   & 4.5   & 5.9   & 3.9   & 5.4   & 5.5 \\
          &       & 0.5   & 5.2   & 4.9   & 4.8   & 5.5   & 5.6   & 4.2   & 6 \\
          & 0.5   & 0     & 6.5   & 6.6   & 6.3   & 7.5   & 4.5   & 6.5   & 6.9 \\
          &       & 0.5   & 6.3   & 6.2   & 6.2   & 6.8   & 7.1   & 5.4   & 6 \\
          & 0.8   & 0     & 8.5   & 7.3   & 8.4   & 8.7   & 9.1   & 9.3   & 6.6 \\
          &       & 0.5   & 10.4  & 7.8   & 8.7   & 6.9   & 8.7   & 8.2   & 8.1 \\
    2000  & 0     & 0     & 5.1   & 5.9   & 5.2   & 4.5   & 4.7   & 5.7   & 4.8 \\
          &       & 0.5   & 3.8   & 5.6   & 5.2   & 5.2   & 5.8   & 5.2   & 6.5 \\
          & 0.5   & 0     & 6.1   & 6.9   & 6     & 6.4   & 5.5   & 6.5   & 5.5 \\
          &       & 0.5   & 5.1   & 5.5   & 5.9   & 6.9   & 7     & 6.6   & 6.4 \\
          & 0.8   & 0     & 9     & 7.4   & 8     & 9.3   & 7.2   & 9.3   & 7.8 \\
          &       & 0.5   & 7.5   & 6.1   & 6.3   & 6.9   & 9.1   & 7.6   & 6.6 \\
    3000  & 0     & 0     & 3.5   & 4.8   & 5.3   & 5.7   & 5     & 5.3   & 5 \\
          &       & 0.5   & 4.3   & 5.2   & 4.4   & 4.2   & 4.4   & 5.1   & 4.5 \\
          & 0.5   & 0     & 4.6   & 5.5   & 6.6   & 6.1   & 6.4   & 6.5   & 6.2 \\
          &       & 0.5   & 5     & 6.4   & 5.3   & 5.1   & 5.1   & 6.1   & 6.3 \\
          & 0.8   & 0     & 7.6   & 6.2   & 6.3   & 6     & 7.2   & 5.9   & 7.4 \\
          &       & 0.5   & 6.4   & 5.3   & 6.8   & 5     & 7.1   & 7     & 6.2 \\
    5000  & 0     & 0     & 5.3   & 6.5   & 4.3   & 4.8   & 6     & 2.9   & 5.5 \\
          &       & 0.5   & 5.2   & 5     & 5.6   & 5.2   & 5.1   & 5.2   & 5 \\
          & 0.5   & 0     & 6.7   & 6.9   & 5.1   & 6     & 7.3   & 3.7   & 5.8 \\
          &       & 0.5   & 6.4   & 5.3   & 6     & 6.3   & 6.2   & 5.5   & 5.7 \\
          & 0.8   & 0     & 6.2   & 5.4   & 7.1   & 6.1   & 6.3   & 6.3   & 5.8 \\
          &       & 0.5   & 5.4   & 6.2   & 7.1   & 4.3   & 7.1   & 5.8   & 6.5 \\
        \bottomrule
		\end{tabular}%
  \label{tab:linearlong}%
	  	  \end{scriptsize}
\end{table}%

\subsection{Polynomial cointegrating regression}\label{subsec:MCpoly}

In this subsection, we consider the case of polynomial cointegrating regression, in particular a quadratic and a cubic relation. We replace the linear model \eqref{eq:linmodel1} and simulate according to
\begin{equation*}
y_t=x_t+x_t^2+u_t,
\end{equation*}
for the quadratic relation, while \eqref{eq:linmodel2} -- \eqref{eq:linmodel4} and all further parameter constellations of Subsection \ref{subsec:MClinear} still hold. We now estimate $\theta=(\theta_1,\theta_2)^{\prime}$ by regressing $y_t$ on $g(x_t,\theta)=\theta_1x_t+\theta_2x_t^2$. In this model, we cannot use the critical values of \cite{Shin1994} as we would treat both $x_t$ and $x_t^2$ as I(1) regressors, see also \cite{Wagner2016}.

Table \ref{tab:quadratic} shows the DNLS test's rejection frequencies. Similar interpretations like in Subsection \ref{subsec:MClinear} for the linear case apply here, too. In addition, we observe a decrease of empirical power relative to Table \ref{tab:linear}, plausibly due to the more complex model to be fitted. 

\begin{table}[tbp]
    \begin{scriptsize}
		\caption{Empirical rejection frequencies for testing the null of cointegration in the \emph{quadratic} regression model for various parameter constellations for the DNLS bootstrap test. All rejection rates are given as percentages. The nominal size is 5\%.}
  \hspace{-0.5cm}
    \begin{tabular}{crrrrrrrrrrrrrrrrr}
    	    \toprule
		& &       & \multicolumn{1}{r}{$T$} & \multicolumn{7}{c}{100}       & \multicolumn{7}{c}{300}    \\
		\cmidrule(l{.75em}){5-11}\cmidrule(l{.75em}){12-18}
     &     &       & \multicolumn{1}{r}{$\tau$} & 0     & \multicolumn{2}{c}{0.1}          & \multicolumn{2}{c}{0.5}   &        \multicolumn{2}{c}{0.9}    & 0     & \multicolumn{2}{c}{0.1}   &  \multicolumn{2}{c}{0.5}  &   \multicolumn{2}{c}{0.9}     \\
					\cmidrule(l{.75em}){6-7}\cmidrule(l{.75em}){8-9}\cmidrule(l{.75em}){10-11}\cmidrule(l{.75em}){13-14}\cmidrule(l{.75em}){15-16}\cmidrule(l{.75em}){17-18}
     &     &       & \multicolumn{1}{r}{$\sigma_1^2$} &       & 1/16 & 16    & 1/16 & 16    & 1/16 & 16    &       & 1/16 & 16    & 1/16 & 16    & 1/16 & 16 \\
  &  \multicolumn{1}{r}{$\rho_{\mu}^2$} & \multicolumn{1}{r}{$\rho$} & \multicolumn{1}{r}{$\lambda$} &       &       &       &       &       &       &       &       &       &       &       &       &       &  \\ \midrule			
          & 0     & 0     & 0     & 5.3   & 8.3   & 4.5   & 6.7   & 4.8   & 4.8   & 3.6   & 7     & 6.6   & 5.9   & 5.1   & 4.7   & 4.9   & 4.6 \\
          &       &       & 0.5   & 4.4   & 7.6   & 4.5   & 5.1   & 5.6   & 5     & 4.1   & 6.6   & 5.5   & 4.2   & 5.1   & 4.4   & 4.1   & 5.4 \\
          &       & 0.5   & 0     & 8.1   & 9.2   & 6.6   & 6.6   & 5.3   & 7.1   & 4.9   & 6.1   & 6.2   & 7.9   & 7     & 5.5   & 6.4   & 5.8 \\
          &       &       & 0.5   & 7.8   & 9.7   & 5.9   & 8.2   & 5.7   & 8.2   & 5.3   & 7.7   & 5.6   & 6.7   & 7.2   & 6.3   & 6.3   & 5.5 \\
          &       & 0.8   & 0     & 13.5  & 14.1  & 11.7  & 11.8  & 9.9   & 14.3  & 12.1  & 11.4  & 8.3   & 9.1   & 9.6   & 7.5   & 10.6  & 7.9 \\\vspace{0.15cm}
          &       &       & 0.5   & 13.6  & 15.6  & 14.4  & 12.8  & 11.5  & 15.5  & 11.7  & 11.1  & 9.3   & 12    & 9.6   & 9.7   & 10.4  & 8.5 \\
          & 0.001 & 0     & 0     & 13    & 19    & 9.7   & 11.1  & 7.2   & 10.5  & 9     & 43.2  & 34.1  & 40.9  & 35.1  & 36.1  & 44.7  & 31.2 \\
          &       &       & 0.5   & 12.8  & 18.3  & 13.3  & 14.5  & 10.5  & 15.4  & 12.7  & 50.9  & 39.1  & 45.4  & 42.2  & 40.8  & 48.1  & 37.7 \\
          &       & 0.5   & 0     & 9.6   & 14.7  & 8     & 9.2   & 5.9   & 9.3   & 6.6   & 21.4  & 21.5  & 18.7  & 19.5  & 18.3  & 21.4  & 15.7 \\
          &       &       & 0.5   & 9.2   & 13.6  & 9.9   & 9.5   & 6.5   & 10.4  & 7.7   & 25.1  & 22.4  & 22.4  & 21.8  & 16.5  & 25.7  & 19.2 \\
          &       & 0.8   & 0     & 13.7  & 17.5  & 11.4  & 12.8  & 11.1  & 15    & 13.1  & 12    & 12.7  & 12.7  & 13.2  & 12    & 14.6  & 14.4 \\
          &       &       & 0.5   & 16.1  & 19.1  & 14.5  & 14.6  & 10.2  & 16.1  & 14.5  & 16.5  & 15.2  & 12.2  & 13    & 12.1  & 14.4  & 13.1 \\
          & 0.01  & 0     & 0     & 44.7  & 44.1  & 40.8  & 36    & 34.3  & 46    & 33.3  & 85.5  & 71.7  & 82.6  & 75.3  & 81.6  & 85.2  & 75.2 \\
          &       &       & 0.5   & 49.1  & 47.6  & 49.3  & 41.5  & 39.1  & 51.5  & 36.9  & 88.2  & 75    & 87.2  & 80.2  & 83    & 87.5  & 80.7 \\
          &       & 0.5   & 0     & 21.7  & 25.9  & 19.7  & 18.3  & 12.7  & 22.4  & 18.2  & 50.5  & 42.9  & 48.2  & 45.1  & 42.7  & 49.4  & 43.4 \\
          &       &       & 0.5   & 23.4  & 27.7  & 20.7  & 21.5  & 15.3  & 22.6  & 20.3  & 52.6  & 45.1  & 49.5  & 44.9  & 43.1  & 52.8  & 46.5 \\
          &       & 0.8   & 0     & 17.9  & 24.2  & 18.4  & 16.8  & 12.8  & 16.3  & 17.4  & 30.6  & 26.8  & 27.7  & 25.5  & 24.8  & 31    & 28.3 \\
          &       &       & 0.5   & 18.1  & 23.4  & 19    & 20.5  & 17.2  & 20.1  & 17.5  & 34.4  & 30.6  & 30    & 32.4  & 28.7  & 34.2  & 27.9 \\
          & 0.1   & 0     & 0     & 81.5  & 77.3  & 78.8  & 75.1  & 74.1  & 81.7  & 72.4  & 98.1  & 95.9  & 97.3  & 95.7  & 97.1  & 97.9  & 97 \\
          &       &       & 0.5   & 83.8  & 82.7  & 80.2  & 75.1  & 77.9  & 83.9  & 77.5  & 99.1  & 95.6  & 98.8  & 97    & 97.6  & 98.8  & 97.7 \\
          &       & 0.5   & 0     & 43.5  & 44.8  & 37.2  & 36.9  & 34.4  & 44    & 41    & 62.6  & 58.2  & 60.9  & 57.2  & 54.7  & 61.1  & 59.4 \\
          &       &       & 0.5   & 47.4  & 45.3  & 39.2  & 37.6  & 35.3  & 45.1  & 42    & 62.9  & 57.6  & 61.8  & 58.1  & 52.1  & 63.5  & 57.7 \\
          &       & 0.8   & 0     & 32.8  & 39.7  & 29.5  & 30.5  & 26.2  & 34.1  & 30.4  & 54.4  & 49.3  & 53.6  & 49.4  & 50.8  & 59.1  & 53.7 \\
          &       &       & 0.5   & 34.6  & 40.9  & 29.3  & 33.4  & 28.1  & 38.3  & 31    & 55.7  & 47.9  & 52.6  & 51.4  & 47.3  & 56.7  & 50.7 \\
        \bottomrule
		\end{tabular}%
  \label{tab:quadratic}%
	  \end{scriptsize}
\end{table}%

Inspired by the application in Section \ref{sec:application}, we also consider a cubic cointegrating regression. We simulate from the model
\begin{equation*}
y_t=1+x_t+2x_t^2+x_t^3+u_t,
\end{equation*}
where the remaining parameters are specified like in the linear and quadratic case. Table \ref{tab:cubicint} shows, analogously to the previous results, the rejection frequencies of the DNLS bootstrap test. We observe that, for the sample sizes considered here and in the presence of endogeneity and autocorrelation, the DNLS test is somewhat oversized with a rejection rate of about 10\%.

\begin{table}[tbp]
   \begin{scriptsize}
		\caption{Empirical rejection frequencies for testing the null of cointegration in the \emph{cubic} regression model for various parameter constellations for the DNLS bootstrap test. All rejection rates are given as percentages. The nominal size is 5\%.
		}
  \hspace{-0.5cm}
    \begin{tabular}{crrrrrrrrrrrrrrrrr}
    	    \toprule
		& &       & \multicolumn{1}{r}{$T$} & \multicolumn{7}{c}{100}       & \multicolumn{7}{c}{300}    \\
		\cmidrule(l{.75em}){5-11}\cmidrule(l{.75em}){12-18}
     &     &       & \multicolumn{1}{r}{$\tau$} & 0     & \multicolumn{2}{c}{0.1}          & \multicolumn{2}{c}{0.5}   &        \multicolumn{2}{c}{0.9}    & 0     & \multicolumn{2}{c}{0.1}   &  \multicolumn{2}{c}{0.5}  &   \multicolumn{2}{c}{0.9}     \\
					\cmidrule(l{.75em}){6-7}\cmidrule(l{.75em}){8-9}\cmidrule(l{.75em}){10-11}\cmidrule(l{.75em}){13-14}\cmidrule(l{.75em}){15-16}\cmidrule(l{.75em}){17-18}
     &     &       & \multicolumn{1}{r}{$\sigma_1^2$} &       & 1/16 & 16    & 1/16 & 16    & 1/16 & 16    &       & 1/16 & 16    & 1/16 & 16    & 1/16 & 16 \\
  &  \multicolumn{1}{r}{$\rho_{\mu}^2$} & \multicolumn{1}{r}{$\rho$} & \multicolumn{1}{r}{$\lambda$} &       &       &       &       &       &       &       &       &       &       &       &       &       &  \\ \midrule		
					& 0     & 0     & 0     & 4     & 6.8   & 3.9   & 4.6   & 6.3   & 4.4   & 7     & 5     & 5.2   & 5     & 3.7   & 5.4   & 5.3   & 5.9 \\
          &       &       & 0.5   & 5.4   & 6.2   & 4.9   & 4.3   & 5.3   & 5.2   & 7.1   & 4.9   & 4.9   & 5.3   & 4.5   & 5.9   & 5.3   & 4.7 \\
          &       & 0.5   & 0     & 6.4   & 8.4   & 7.8   & 5.4   & 7.6   & 7     & 7.8   & 8.3   & 8.3   & 8.3   & 6     & 7.1   & 8.8   & 7 \\
          &       &       & 0.5   & 8.4   & 9.4   & 8.3   & 7.4   & 7.8   & 9.1   & 8.8   & 7.9   & 7.7   & 6.3   & 6.7   & 8     & 7.5   & 8.6 \\
          &       & 0.8   & 0     & 15.9  & 18.1  & 15    & 12.4  & 13.2  & 15.4  & 14.2  & 14.6  & 10.4  & 12.1  & 12.1  & 9.1   & 12.7  & 9.3 \\\vspace{0.15cm}
          &       &       & 0.5   & 19.2  & 21.4  & 18.2  & 16.7  & 15    & 20    & 16.3  & 15.1  & 12    & 16.1  & 13.3  & 12.3  & 14.9  & 12.1 \\
          & 0.001 & 0     & 0     & 9.9   & 10.3  & 10.6  & 8.5   & 7.8   & 10.7  & 10.2  & 45.6  & 25.1  & 45.5  & 28.7  & 31    & 44.6  & 21.2 \\
          &       &       & 0.5   & 12.6  & 10.4  & 13.9  & 8.8   & 8.2   & 14    & 11.3  & 49.1  & 28.9  & 49.1  & 35.8  & 37.5  & 50    & 29 \\
          &       & 0.5   & 0     & 7.6   & 9.7   & 8.2   & 7.6   & 6.5   & 10    & 8.7   & 23.6  & 14.7  & 24.4  & 14.8  & 17.2  & 25.1  & 12.7 \\
          &       &       & 0.5   & 9.6   & 10.4  & 10.5  & 6.8   & 7     & 10.2  & 11.4  & 26.2  & 15.6  & 24.6  & 18.3  & 18    & 26.3  & 17.2 \\
          &       & 0.8   & 0     & 15.6  & 17.5  & 14.1  & 14.8  & 11.4  & 15.6  & 13    & 16.2  & 11    & 15.7  & 11.9  & 12.7  & 15.2  & 11.5 \\
          &       &       & 0.5   & 17.6  & 21.7  & 19.9  & 18.4  & 12.7  & 20.2  & 17.6  & 17.8  & 13.3  & 17.7  & 14.1  & 14.3  & 17.2  & 14.6 \\
          & 0.01  & 0     & 0     & 41.9  & 31.4  & 45.7  & 30.6  & 31.7  & 44.2  & 29.3  & 89.6  & 73.9  & 89.4  & 79.1  & 80    & 90.4  & 71.7 \\
          &       &       & 0.5   & 47.2  & 35.9  & 52.4  & 32.5  & 39    & 50.4  & 30.9  & 91.3  & 79    & 92.5  & 85.5  & 83.6  & 90.4  & 77.3 \\
          &       & 0.5   & 0     & 20.4  & 16.6  & 22.5  & 14.9  & 17.1  & 22.4  & 16.9  & 57.8  & 40.9  & 60.2  & 46.7  & 46.2  & 61.8  & 40 \\
          &       &       & 0.5   & 24.6  & 19.3  & 26.7  & 15.6  & 18.2  & 27.4  & 16.8  & 61    & 47.7  & 64    & 50.7  & 48.3  & 63.8  & 45.1 \\
          &       & 0.8   & 0     & 19.3  & 20.6  & 20.6  & 16.6  & 15.6  & 23.7  & 15.4  & 31.8  & 23.2  & 35    & 27.9  & 29    & 37.5  & 23.1 \\
          &       &       & 0.5   & 21.8  & 24.9  & 24.6  & 19.2  & 16.1  & 23.2  & 20.1  & 36.9  & 24.1  & 40.8  & 29.2  & 31    & 40    & 26.6 \\
          & 0.1   & 0     & 0     & 84.7  & 74.1  & 88    & 71.9  & 75    & 86.7  & 70.8  & 99.5  & 98.2  & 99.4  & 98.4  & 98.7  & 99.7  & 97.5 \\
          &       &       & 0.5   & 88.9  & 78.8  & 89.6  & 78.4  & 83.2  & 89.5  & 76.7  & 99.7  & 97.7  & 99.9  & 99.5  & 99.4  & 99.9  & 98.7 \\
          &       & 0.5   & 0     & 49.7  & 40    & 52.2  & 33.5  & 38.3  & 52.5  & 38    & 77.2  & 63.2  & 79.1  & 69.6  & 69.9  & 77.3  & 64 \\
          &       &       & 0.5   & 52.4  & 42.8  & 50.7  & 38.2  & 43.9  & 52.8  & 42.9  & 78.5  & 65.6  & 79.9  & 70.8  & 68.9  & 81    & 63.7 \\
          &       & 0.8   & 0     & 36.2  & 33.7  & 37    & 27.1  & 26.1  & 38.2  & 29.5  & 65.3  & 49.9  & 68.1  & 53.4  & 55.8  & 69.7  & 53.2 \\
          &       &       & 0.5   & 44.6  & 36.9  & 43.4  & 28.1  & 31    & 42.7  & 34.4  & 67.5  & 50.6  & 72.2  & 56.1  & 57    & 70.4  & 53.4 \\          
       \bottomrule
    \end{tabular}%
  \label{tab:cubicint}%
		\end{scriptsize}
\end{table}%

\subsection{Smooth transition regression model}\label{subsec:smooth}

We now discuss an example of a cointegrating regression which is indeed nonlinear in the parameters. Thus, NLS is needed for (first-step) estimation. We adopt the example of cointegrating smooth transition functions also considered in \cite{SaikkonenChoi2004} and \cite{ChoiSaikkonen2010} (augmented with heteroskedasticity). We generate data according to
\begin{equation*}
y_t=\delta_0+\theta_1x_t+\theta_2\frac{1}{1+\exp(-(x_t-\theta_3))}+u_t,
\end{equation*}
with the parameter constellation $\delta_0=0,\theta_1=1,\theta_2=1,\theta_3=5$. The value of $\theta_3$, in this DGP, dictates the location at which the relationship between regressor and regressand changes. See Figure 1 in \citet{SaikkonenChoi2004}, with their $c=\theta_3$. Hence, $\theta_3$ acts, effectively, like a location parameter shifting the ``point of nonlinearity''.\footnote{Here, we assume that the transition variable is non-stationary because of our assumption that $x_t$ is $I(1)$. The analysis of a smooth cointegrating regression with a stationary transition variable and, more generally, mixtures of $I(1)$ and $I(0)$ would need a relaxation of the assumptions. An $I(0)$ transition variable would imply, at least, a particular functional form of $g$, for example one that is such that the dependence on $x_t\sim I(1)$ is such that it is filtered into an $I(0)$ variable.} 

Table \ref{tab:smoothint} reports the rejection rates for the DNLS bootstrap test.\footnote{We reduced the number of bootstrap replication $B$ to 200 for this case as the nonlinear bootstrap simulations are very computationally demanding. In rare cases, for some generated samples the NLS algorithm does not converge. We thus exclude these cases from the analysis.} We observe that the bootstrap test works reasonably well, again with some moderate size problems in the presence of either endogeneity or autocorrelation and somewhat larger size distortions for both endogeneity and autocorrelation.

\begin{table}[tbp]
  \begin{scriptsize}
		\caption{Empirical rejection frequencies for testing the null of cointegration in the \emph{smooth transition} regression model for various parameter constellations for the DNLS bootstrap test. All rejection rates are given as percentages. The nominal size is 5\%.
		}
  \hspace{-0.5cm}
    \begin{tabular}{crrrrrrrrrrrrrrrrr}
    	    \toprule
		& &       & \multicolumn{1}{r}{$T$} & \multicolumn{7}{c}{100}       & \multicolumn{7}{c}{300}    \\
		\cmidrule(l{.75em}){5-11}\cmidrule(l{.75em}){12-18}
     &     &       & \multicolumn{1}{r}{$\tau$} & 0     & \multicolumn{2}{c}{0.1}          & \multicolumn{2}{c}{0.5}   &        \multicolumn{2}{c}{0.9}    & 0     & \multicolumn{2}{c}{0.1}   &  \multicolumn{2}{c}{0.5}  &   \multicolumn{2}{c}{0.9}     \\
					\cmidrule(l{.75em}){6-7}\cmidrule(l{.75em}){8-9}\cmidrule(l{.75em}){10-11}\cmidrule(l{.75em}){13-14}\cmidrule(l{.75em}){15-16}\cmidrule(l{.75em}){17-18}
     &     &       & \multicolumn{1}{r}{$\sigma_1^2$} &       & 1/16 & 16    & 1/16 & 16    & 1/16 & 16    &       & 1/16 & 16    & 1/16 & 16    & 1/16 & 16 \\
  &  \multicolumn{1}{r}{$\rho_{\mu}^2$} & \multicolumn{1}{r}{$\rho$} & \multicolumn{1}{r}{$\lambda$} &       &       &       &       &       &       &       &       &       &       &       &       &       &  \\ \midrule		
					& 0     & 0     & 0     & 4.6   & 7.1   & 4.6   & 5.8   & 4.4   & 5.2   & 8.7   & 6     & 6.7   & 3.9   & 4.1   & 5.2   & 4.5   & 6 \\
          &       &       & 0.5   & 5.4   & 6.9   & 5     & 5.7   & 5.3   & 5.6   & 7.2   & 5.2   & 6.9   & 5.6   & 4.5   & 6.4   & 5.2   & 5.4 \\
          &       & 0.5   & 0     & 7.9   & 10.5  & 7     & 7.1   & 6     & 8.8   & 10.1  & 6.8   & 7     & 6.9   & 6.6   & 7     & 7.8   & 7.2 \\
          &       &       & 0.5   & 8     & 8.7   & 9.5   & 6.8   & 6.4   & 7.5   & 7.4   & 8     & 7     & 7     & 7.2   & 8.7   & 8.2   & 6.3 \\
          &       & 0.8   & 0     & 17    & 19.2  & 16.9  & 10.6  & 11.8  & 16.2  & 14.7  & 11.9  & 9.2   & 11.1  & 10.7  & 9.3   & 11.8  & 11.6 \\\vspace{0.15cm}
          &       &       & 0.5   & 20.6  & 22.2  & 19.1  & 18.9  & 12.1  & 18.9  & 17.3  & 15.2  & 12.7  & 14.9  & 12.3  & 12.2  & 16.9  & 11.1 \\
          & 0.001 & 0     & 0     & 11    & 12    & 11.5  & 7.6   & 6.6   & 10.6  & 9.7   & 45.1  & 23    & 45.6  & 34.8  & 32.2  & 47.9  & 23.3 \\
          &       &       & 0.5   & 12.5  & 14.1  & 14    & 11    & 8.7   & 12.3  & 8.2   & 52.2  & 30.8  & 51.5  & 35.5  & 38.9  & 54.1  & 29.4 \\
          &       & 0.5   & 0     & 10.3  & 9.7   & 8.9   & 6.1   & 7.5   & 8.6   & 9.4   & 23.4  & 12.7  & 23    & 16.4  & 16    & 25    & 12.9 \\
          &       &       & 0.5   & 11.2  & 11.8  & 11.6  & 8.4   & 7.2   & 9     & 8.4   & 25.7  & 17.1  & 26.7  & 20.1  & 19.3  & 26.6  & 16.2 \\
          &       & 0.8   & 0     & 17.5  & 17    & 16.6  & 12.3  & 13.1  & 16.9  & 13.2  & 15.1  & 12.7  & 17.1  & 13.5  & 13.2  & 17.7  & 10.9 \\
          &       &       & 0.5   & 23.7  & 19.9  & 18.8  & 19.7  & 13.6  & 23.1  & 17    & 19.1  & 14.2  & 19.2  & 14.9  & 13.9  & 18.3  & 14.8 \\
          & 0.01  & 0     & 0     & 45.3  & 30.6  & 44.9  & 29.4  & 33.8  & 42.1  & 26.2  & 91.2  & 72.9  & 90    & 79    & 81.3  & 91.7  & 75.2 \\
          &       &       & 0.5   & 49.2  & 36.2  & 50.8  & 32.4  & 37.9  & 51    & 36.6  & 94.2  & 75.9  & 93.2  & 85.3  & 85.6  & 92.3  & 78.6 \\
          &       & 0.5   & 0     & 24.3  & 17.5  & 23.2  & 14.9  & 15.7  & 19.5  & 15.4  & 60.4  & 43.5  & 61.4  & 48.6  & 46.2  & 62.2  & 41.8 \\
          &       &       & 0.5   & 24.3  & 19.1  & 27.4  & 15.8  & 17.8  & 28    & 19.3  & 63.8  & 44.1  & 62.6  & 54.1  & 49.9  & 64.7  & 44.8 \\
          &       & 0.8   & 0     & 21.7  & 20.5  & 20.6  & 15.6  & 13.5  & 20.2  & 17.6  & 35.8  & 25.9  & 36    & 26.9  & 23.4  & 36.7  & 22.5 \\
          &       &       & 0.5   & 27.4  & 24.2  & 25.6  & 19    & 18.5  & 24.3  & 20    & 38.8  & 28.2  & 39    & 30.9  & 27.6  & 39.2  & 27.3 \\
          & 0.1   & 0     & 0     & 84.4  & 74.3  & 87.4  & 73.1  & 74.4  & 86.2  & 70.9  & 99.9  & 97.4  & 99.8  & 98.3  & 99.2  & 99.8  & 97.7 \\
          &       &       & 0.5   & 91.7  & 77.9  & 91.4  & 77.3  & 81.7  & 90    & 75.4  & 99.7  & 97.9  & 99.8  & 99.2  & 98.9  & 99.8  & 98.4 \\
          &       & 0.5   & 0     & 48.5  & 41.3  & 53.7  & 36.4  & 34    & 52.7  & 37.2  & 79.9  & 61.7  & 78.2  & 67.9  & 70.2  & 82.7  & 63.7 \\
          &       &       & 0.5   & 57.4  & 40.9  & 54.8  & 37.2  & 39.5  & 54.5  & 41.6  & 76.1  & 61.9  & 81.7  & 68.7  & 70    & 79.8  & 64.2 \\
          &       & 0.8   & 0     & 38.1  & 31.1  & 39.5  & 24.5  & 27.3  & 39.6  & 27.7  & 69.2  & 53    & 73.6  & 59.2  & 54.3  & 69.5  & 52.3 \\
          &       &       & 0.5   & 44    & 33.2  & 47.3  & 31.9  & 29    & 39.7  & 33.6  & 70    & 53.9  & 72.9  & 56    & 56.7  & 70.5  & 54.3 \\
    \bottomrule
    \end{tabular}%
  \label{tab:smoothint}%
		\end{scriptsize}
\end{table}%

\subsection{Threshold regression model}\label{subsec:TAR}

We consider a threshold cointegrating regression model which was proposed by \cite{Gonzalo2006}. We generate data according to
\begin{equation*}
y_t=\theta_1x_t+\theta_2x_t\textbf{1}(q_{t-r}>\theta_3)+u_t,
\end{equation*}
with the parameter constellation $\theta_1=1,\theta_2=0.15,\theta_3=0$. The threshold variable $q_{t-r}$ is a stationary process lagged by $r\ge1$ periods. Here we specify an AR(1) process $q_t=0.5 q_{t-1}+\epsilon_t$, with $\epsilon_t\sim N(0,1)$ i.i.d., and set $r:=1$. To estimate $(\theta_1,\theta_2,\theta_3)$ we first consider $\theta_3$ as fixed and run OLS to estimate the remaining parameters. We then estimate $\theta_3$ by minimizing the sum of squared residuals of the OLS estimations. Since this estimation scheme is computationally demanding we abstain from using DNLS for this case and only present the bootstrap test based on residuals of the NLS estimation described above.

Table \ref{tab:TAR} presents empirical size and power of the OLS bootstrap test. We observe a qualitatively similar picture as for the other model constellations.

\begin{table}[tbp]
    \begin{scriptsize}
		\caption{Empirical rejection frequencies for testing the null of cointegration in the \emph{threshold} regression model for various parameter constellations for the bootstrap test using OLS. All rejection rates are given as percentages. The nominal size is 5\%. }
  \hspace{-0.5cm}
    \begin{tabular}{crrrrrrrrrrrrrrrrr}
    	    \toprule
		& &       & \multicolumn{1}{r}{$T$} & \multicolumn{7}{c}{100}       & \multicolumn{7}{c}{300}    \\
		\cmidrule(l{.75em}){5-11}\cmidrule(l{.75em}){12-18}
     &     &       & \multicolumn{1}{r}{$\tau$} & 0     & \multicolumn{2}{c}{0.1}          & \multicolumn{2}{c}{0.5}   &        \multicolumn{2}{c}{0.9}    & 0     & \multicolumn{2}{c}{0.1}   &  \multicolumn{2}{c}{0.5}  &   \multicolumn{2}{c}{0.9}     \\
					\cmidrule(l{.75em}){6-7}\cmidrule(l{.75em}){8-9}\cmidrule(l{.75em}){10-11}\cmidrule(l{.75em}){13-14}\cmidrule(l{.75em}){15-16}\cmidrule(l{.75em}){17-18}
     &     &       & \multicolumn{1}{r}{$\sigma_1^2$} &       & 1/16 & 16    & 1/16 & 16    & 1/16 & 16    &       & 1/16 & 16    & 1/16 & 16    & 1/16 & 16 \\
  &  \multicolumn{1}{r}{$\rho_{\mu}^2$} & \multicolumn{1}{r}{$\rho$} & \multicolumn{1}{r}{$\lambda$} &       &       &       &       &       &       &       &       &       &       &       &       &       &  \\ \midrule
					& 0     & 0     & 0     & 5.5   & 7.7   & 4.8   & 5.5   & 4.7   & 4.4   & 4.8   & 4.7   & 5.6   & 5.8   & 4.8   & 5     & 6.6   & 5.4 \\
          &       &       & 0.5   & 4.9   & 6.7   & 3.8   & 5     & 3.9   & 4.3   & 6.2   & 3.2   & 5.3   & 3.5   & 4.9   & 4.1   & 4.8   & 6.3 \\
          &       & 0.5   & 0     & 7.9   & 6.2   & 7.5   & 8     & 5.4   & 7.9   & 6.4   & 7.6   & 7.2   & 8.1   & 7.4   & 6.6   & 8.9   & 6.1 \\
          &       &       & 0.5   & 7.6   & 6.4   & 6.9   & 8.8   & 5.5   & 5.8   & 6.3   & 5.8   & 6.9   & 5.8   & 5.7   & 6.3   & 6.7   & 8.3 \\
          &       & 0.8   & 0     & 12.4  & 11.1  & 11.4  & 14.5  & 12    & 11.2  & 13.6  & 10.5  & 8.9   & 10.1  & 11.7  & 10.6  & 10.3  & 10.3 \\\vspace{0.15cm}
          &       &       & 0.5   & 11.6  & 13.8  & 11.9  & 13    & 8.6   & 12.3  & 12    & 10.8  & 9.3   & 8.5   & 11.5  & 9.4   & 10.3  & 12.4 \\
          & 0.001 & 0     & 0     & 14.6  & 14.8  & 12.9  & 14.6  & 10.4  & 15.3  & 10.2  & 49.2  & 35.2  & 46.3  & 45    & 42.9  & 45.9  & 37.5 \\
          &       &       & 0.5   & 14.1  & 17    & 12.8  & 13.1  & 10.3  & 14.8  & 11.6  & 50.3  & 39.8  & 45.2  & 43.7  & 42.1  & 49.8  & 39 \\
          &       & 0.5   & 0     & 11.7  & 10    & 10.3  & 10.9  & 7.1   & 9.9   & 8.1   & 26.8  & 23.9  & 22.8  & 24.7  & 21.5  & 23.9  & 21.1 \\
          &       &       & 0.5   & 9.4   & 12.7  & 9     & 10.7  & 6.5   & 10.8  & 8.5   & 28.8  & 24.6  & 20.6  & 24.3  & 19.6  & 25.5  & 21.6 \\
          &       & 0.8   & 0     & 15.8  & 12.9  & 13.7  & 14.5  & 11.9  & 14.6  & 11.5  & 15.5  & 16.3  & 15    & 17.1  & 13.4  & 17.2  & 13.1 \\
          &       &       & 0.5   & 14.9  & 13.7  & 12.6  & 14.9  & 9.5   & 12.1  & 13.6  & 13.7  & 16    & 15.3  & 14.3  & 12.4  & 15.2  & 15.7 \\
          & 0.01  & 0     & 0     & 45.6  & 43.2  & 45.2  & 43.5  & 41.5  & 51.2  & 39.9  & 88.1  & 74    & 84.2  & 79.4  & 86.3  & 85.7  & 80.9 \\
          &       &       & 0.5   & 46.1  & 40.4  & 45.6  & 42.1  & 41.2  & 48    & 38.7  & 87    & 77.7  & 84.1  & 80.3  & 86.7  & 87.4  & 78.5 \\
          &       & 0.5   & 0     & 22.2  & 26.1  & 20.4  & 22.6  & 15.3  & 24.9  & 20.4  & 54.5  & 47    & 52.1  & 49.2  & 52.1  & 56.9  & 51.6 \\
          &       &       & 0.5   & 20.7  & 24.8  & 19.7  & 22.3  & 16.2  & 21.9  & 20.8  & 55    & 48.3  & 52.4  & 50.1  & 50    & 56.3  & 46.7 \\
          &       & 0.8   & 0     & 19.8  & 21.4  & 18.2  & 19.2  & 13.6  & 20.1  & 17    & 35.4  & 35.3  & 35.2  & 33.9  & 31    & 38.4  & 30.6 \\
          &       &       & 0.5   & 16.1  & 22.3  & 19.3  & 21.1  & 15.6  & 20.4  & 17.7  & 35.7  & 35.4  & 34.1  & 32.8  & 30    & 37.4  & 33.5 \\
          & 0.1   & 0     & 0     & 84.6  & 77.4  & 77.8  & 76.2  & 79.1  & 83.4  & 77    & 98.5  & 96.1  & 97.9  & 95.6  & 98.8  & 98.7  & 97.3 \\
          &       &       & 0.5   & 84.4  & 75    & 81.2  & 77.5  & 79.5  & 83    & 77.1  & 98.9  & 97    & 98.3  & 95.8  & 98.5  & 99    & 97.1 \\
          &       & 0.5   & 0     & 44.5  & 47.4  & 40.9  & 40.9  & 35.9  & 47    & 42.1  & 65.6  & 65.8  & 62.2  & 64.2  & 64.4  & 69.2  & 63.7 \\
          &       &       & 0.5   & 43    & 44.1  & 41.8  & 42.3  & 34.3  & 43.8  & 40    & 69.1  & 62.9  & 65    & 64.7  & 62.5  & 69.9  & 62 \\
          &       & 0.8   & 0     & 35    & 36.2  & 34.8  & 33.3  & 27.2  & 33.7  & 28.2  & 61    & 58.5  & 61    & 58    & 57.2  & 63.8  & 55.1 \\
          &       &       & 0.5   & 37    & 38.6  & 32.7  & 31.6  & 29.3  & 37.7  & 30.8  & 63.2  & 58.7  & 55.3  & 56.8  & 53.9  & 61.1  & 58.6 \\        
    \bottomrule
    \end{tabular}%
  \label{tab:TAR}%
		\end{scriptsize}
\end{table}%

\subsection{Trend regressors}\label{app:simulationstrend}

We next extend the simulations of Section \ref{subsec:MCpoly} to study the impact a time trend on the rejection rates for the bootstrap test. More specifically, we discuss the cubic regression model with time trend as it is also employed in Section \ref{sec:application}. Coefficients of trend regressors can either be included into the model and estimated or, equivalently (by the Frisch-Waugh-Lovell theorem), regressors and regressand can be de-trended before estimating the original model.

We specify the model
\begin{equation*}
y_t=1+t+x_t+2x_t^2+x_t^3+u_t.
\end{equation*}
All other parameters are chosen as in the other setups. Table \ref{tab:cubicdet} reports rejection rates. We observe a fairly similar picture for the cubic cointegration regression with trend as for the cubic regression without time trend, see Table \ref{tab:cubicint}. The empirical size is highly comparable for both setups. However, and as one would expect from the additional parameter to be fitted, the additional time regressor comes with a reduction of power.

\begin{table}[tbp]
    \begin{scriptsize}
		\caption{Empirical rejection frequencies for testing the null of cointegration in the \emph{cubic} regression model with \emph{time trend} for various parameter constellations for the DNLS bootstrap test. All rejection rates are given as percentages. The nominal size is 5\%.		}
  \hspace{-0.5cm}
    \begin{tabular}{crrrrrrrrrrrrrrrrr}
    	    \toprule
		& &       & \multicolumn{1}{r}{$T$} & \multicolumn{7}{c}{100}       & \multicolumn{7}{c}{300}    \\
		\cmidrule(l{.75em}){5-11}\cmidrule(l{.75em}){12-18}
     &     &       & \multicolumn{1}{r}{$\tau$} & 0     & \multicolumn{2}{c}{0.1}          & \multicolumn{2}{c}{0.5}   &        \multicolumn{2}{c}{0.9}    & 0     & \multicolumn{2}{c}{0.1}   &  \multicolumn{2}{c}{0.5}  &   \multicolumn{2}{c}{0.9}     \\
					\cmidrule(l{.75em}){6-7}\cmidrule(l{.75em}){8-9}\cmidrule(l{.75em}){10-11}\cmidrule(l{.75em}){13-14}\cmidrule(l{.75em}){15-16}\cmidrule(l{.75em}){17-18}
     &     &       & \multicolumn{1}{r}{$\sigma_1^2$} &       & 1/16 & 16    & 1/16 & 16    & 1/16 & 16    &       & 1/16 & 16    & 1/16 & 16    & 1/16 & 16 \\
  &  \multicolumn{1}{r}{$\rho_{\mu}^2$} & \multicolumn{1}{r}{$\rho$} & \multicolumn{1}{r}{$\lambda$} &       &       &       &       &       &       &       &       &       &       &       &       &       &  \\ \midrule
      
			   	& 0     & 0     & 0     & 4.2   & 8.3   & 3.4   & 5.7   & 5.3   & 4.3   & 7.2   & 3.3   & 5.9   & 5     & 5.4   & 6     & 4.1   & 6.8 \\
          &       &       & 0.5   & 6     & 7.5   & 5.6   & 5.2   & 5.2   & 3.5   & 7.8   & 5.1   & 6.2   & 5.6   & 4.2   & 5.1   & 6.4   & 5.5 \\
          &       & 0.5   & 0     & 4.8   & 7.5   & 4.7   & 4.5   & 4.8   & 5.5   & 8.4   & 7.1   & 7.5   & 8.7   & 6.4   & 8.6   & 6.5   & 8.8 \\
          &       &       & 0.5   & 7.9   & 7.9   & 5.6   & 4.7   & 4.8   & 5.5   & 7     & 8.8   & 7.2   & 7.2   & 7     & 7.9   & 9.4   & 8.1 \\
          &       & 0.8   & 0     & 12.4  & 15.5  & 12.6  & 9.4   & 9.2   & 9.8   & 11.8  & 13.2  & 13.2  & 15    & 11.2  & 10    & 13.9  & 9.5 \\\vspace{0.15cm}
          &       &       & 0.5   & 14.2  & 20.2  & 14.5  & 13.7  & 11.7  & 15.5  & 14.4  & 16.4  & 13.8  & 17.1  & 14.6  & 15.9  & 17    & 13.7 \\
          & 0.001 & 0     & 0     & 7.9   & 9.2   & 7.4   & 5.9   & 6.6   & 6.4   & 8.8   & 23.7  & 13.6  & 26.7  & 20    & 20.8  & 25.2  & 14.2 \\
          &       &       & 0.5   & 8     & 8.7   & 7.6   & 9     & 7.3   & 7.8   & 7.7   & 32    & 15.8  & 29.9  & 25.1  & 26    & 30.4  & 14.3 \\
          &       & 0.5   & 0     & 7.2   & 7.5   & 6.9   & 5     & 4.5   & 5.3   & 7.2   & 12.8  & 11.2  & 15.2  & 12.5  & 12    & 15.1  & 9.4 \\
          &       &       & 0.5   & 7.3   & 8.8   & 6.9   & 5.5   & 4.1   & 5.5   & 6.4   & 17    & 11.8  & 14.6  & 12.5  & 12.3  & 13.8  & 10.3 \\
          &       & 0.8   & 0     & 13.3  & 17.8  & 10.6  & 10.4  & 9.8   & 10.6  & 13.6  & 17.3  & 13.5  & 16.3  & 12.7  & 13.6  & 14.4  & 12.5 \\
          &       &       & 0.5   & 13.4  & 20    & 14.7  & 11.3  & 9.9   & 14.5  & 15.6  & 19.9  & 16.3  & 17    & 15    & 16    & 19.6  & 15.7 \\
          & 0.01  & 0     & 0     & 21.3  & 17.9  & 25.2  & 20.9  & 22.1  & 24.1  & 16.7  & 82.2  & 60.6  & 82.3  & 70.7  & 73.4  & 80.7  & 61.8 \\
          &       &       & 0.5   & 26.2  & 22.7  & 29.2  & 23.3  & 25.4  & 31.1  & 20.3  & 86.7  & 67.6  & 88    & 81.1  & 80.7  & 86    & 68.1 \\
          &       & 0.5   & 0     & 9.8   & 10.1  & 9.1   & 9     & 8.6   & 10.1  & 8.1   & 42.4  & 27.7  & 44.5  & 35.3  & 35.9  & 42.3  & 26.4 \\
          &       &       & 0.5   & 10.8  & 11.6  & 11.3  & 8.7   & 7.7   & 12    & 9.4   & 47.2  & 30.5  & 48.6  & 42    & 42.4  & 47.5  & 31.1 \\
          &       & 0.8   & 0     & 14.6  & 17.5  & 15    & 12.2  & 9.5   & 12.5  & 13.4  & 25.7  & 17.4  & 24.8  & 22.1  & 20.4  & 21.7  & 14.8 \\
          &       &       & 0.5   & 14.3  & 19.6  & 16.2  & 14.2  & 13.3  & 15.4  & 16.6  & 29.7  & 20.3  & 29.9  & 24.8  & 23.8  & 28.2  & 22.2 \\
          & 0.1   & 0     & 0     & 76.7  & 62.2  & 77.1  & 67    & 69    & 79.3  & 59.1  & 99.5  & 98.4  & 99.9  & 98.8  & 99.2  & 99.6  & 96.8 \\
          &       &       & 0.5   & 84.6  & 69.7  & 83.6  & 73.3  & 79.1  & 84.8  & 66.3  & 99.8  & 97.9  & 100   & 99.4  & 99.3  & 99.9  & 99.1 \\
          &       & 0.5   & 0     & 29.8  & 21.3  & 27.1  & 22.9  & 23.1  & 30.5  & 19.9  & 71.2  & 59.4  & 74.7  & 65.1  & 67.7  & 72.5  & 54.2 \\
          &       &       & 0.5   & 30    & 23.8  & 28    & 24.6  & 27.6  & 29    & 22.8  & 74.3  & 57.3  & 76.8  & 66.6  & 68.6  & 75.6  & 59.1 \\
          &       & 0.8   & 0     & 18.1  & 24.8  & 19.9  & 17.6  & 17.8  & 18    & 20.1  & 54.1  & 38.7  & 55.6  & 49.2  & 49.6  & 58.2  & 45.1 \\
          &       &       & 0.5   & 25.1  & 23.4  & 23.3  & 21.8  & 17.2  & 22.3  & 22.5  & 56.8  & 42.3  & 60.9  & 51.9  & 54.2  & 59.5  & 46.3 \\
    \bottomrule
    \end{tabular}%
  \label{tab:cubicdet}%
		\end{scriptsize}
\end{table}%

\subsection{Sieve bootstrap}\label{subsec:sieve}

In this subsection, we present an alternative bootstrap approach which is based on the VAR sieve bootstrap principle \cite[]{Chang/ParkETAL:06,Reichold+Jentsch-BootInfeCoinRegr:23}. We build on \cite{Lee2012} who present a sieve bootstrap to improve the standard KPSS test. Our version of the sieve bootstrap to test the null of cointegration proceeds as follows.
\begin{enumerate}
\item Run the original DNLS regression, save residuals $\hat{e}_{Kt}$ from \eqref{eq:LLresiduals} and compute the test statistic $\hat{\eta}_{DNLS}$ as given in \eqref{eq:etaLL}. Define $\hat{w}_{t}=(\hat{e}_{Kt},v_t^{\prime})^{\prime}$ with $v_t=\Delta x_t$.
\item Estimate the $VAR(q)$
\begin{equation*}
\hat{w}_{t}=\hat\Phi_1\hat{w}_{t-1}+\ldots+\hat\Phi_q\hat{w}_{t-q}+\hat\varepsilon_{qt}
\end{equation*}
to obtain the residuals $\hat\varepsilon_{qt}$. The optimal $q$ may be chosen using the AIC.
\item Compute the centered residuals $\tilde\varepsilon_{qt}:=\hat\varepsilon_{qt}-\frac{1}{T-q}\sum_{t=q+1}^T\hat\varepsilon_{qt}$ for $t=q+1,\ldots,T$ times a standard normal r.v., denoted as $\varepsilon_{t}^b$ for $t=1,\ldots,T$; i.e., $\hat\varepsilon_{t}^b:=\tilde\varepsilon_{qt}\varepsilon_{t}^b$.\footnote{We also experimented with a version without the standard normal factor. The size results are qualitatively similar, the power is less. We thus omit the results.}
\item Construct the bootstrap sample recursively using
\begin{equation*}
\hat{w}_{t}^b=\hat\Phi_1\hat{w}_{t-1}^b+\ldots+\hat\Phi_q\hat{w}_{t-q}^b+\hat\varepsilon_{t}^b
\end{equation*}
given initial values $w_{1-q}^b,\ldots,w_0^b$. Partition $w_t^b=(\hat{e}_{t}^b,v_t^{b\prime})^{\prime}$ analogously to $\hat{w}_{t}$ and define $x_t^b:=\sum_{s=1}^tv_s$ and $y_t^b:=h(t,x_t^b,\hat\vartheta)+\hat{e}_{t}^b$.
\item Estimate the DNLS test statistic based on the bootstrap sample.
\item Repeat steps 3 to 5 independently $B$ times and compute the simulated bootstrap $p$-value $1-\tilde{G}_T^b(\hat{\eta}_{DNLS})$, where $\tilde{G}_T^b$ is the empirical cumulative distribution function of the bootstrap test statistics.
\end{enumerate}
Table \ref{tab:sieve} shows empirical rejection rates for the linear DGP of Subsection \ref{subsec:MClinear}. As in panel (b) of Table \ref{tab:linear} for the fixed-regressor bootstrap, we observe that the sieve bootstrap is performing better than the fixed-regressor bootstrap in many scenarios. However, the sieve bootstrap has less power than the fixed-regressor bootstrap for almost all of the scenarios. We conclude that a full theoretical analysis of the sieve would be an interesting avenue for further research.

\begin{table}[tbp]
    \begin{scriptsize}
		\caption{Empirical rejection frequencies for testing the null of cointegration in the \emph{linear} regression model using the VAR sieve bootstrap for various parameter constellations. All rejection rates are given as percentages. The nominal size is 5\%.		}
  \hspace{-0.5cm}
    \begin{tabular}{crrrrrrrrrrrrrrrrr}
    	    \toprule
		& &       & \multicolumn{1}{r}{$T$} & \multicolumn{7}{c}{100}       & \multicolumn{7}{c}{300}    \\
		\cmidrule(l{.75em}){5-11}\cmidrule(l{.75em}){12-18}
     &     &       & \multicolumn{1}{r}{$\tau$} & 0     & \multicolumn{2}{c}{0.1}          & \multicolumn{2}{c}{0.5}   &        \multicolumn{2}{c}{0.9}    & 0     & \multicolumn{2}{c}{0.1}   &  \multicolumn{2}{c}{0.5}  &   \multicolumn{2}{c}{0.9}     \\
					\cmidrule(l{.75em}){6-7}\cmidrule(l{.75em}){8-9}\cmidrule(l{.75em}){10-11}\cmidrule(l{.75em}){13-14}\cmidrule(l{.75em}){15-16}\cmidrule(l{.75em}){17-18}
     &     &       & \multicolumn{1}{r}{$\sigma_1^2$} &       & 1/16 & 16    & 1/16 & 16    & 1/16 & 16    &       & 1/16 & 16    & 1/16 & 16    & 1/16 & 16 \\
  &  \multicolumn{1}{r}{$\rho_{\mu}^2$} & \multicolumn{1}{r}{$\rho$} & \multicolumn{1}{r}{$\lambda$} &       &       &       &       &       &       &       &       &       &       &       &       &       &  \\ \midrule
					
					& 0     & 0     & 0     & 5     & 7.1   & 5     & 4.4   & 4.6   & 5.5   & 4     & 5.6   & 7.1   & 5.3   & 5.5   & 4.9   & 4.8   & 5.1 \\
          &       &       & 0.5   & 4.5   & 9.1   & 5.1   & 4.5   & 4.4   & 4.6   & 4.9   & 4.4   & 6.1   & 4.9   & 6.9   & 3     & 3.7   & 5.9 \\
          &       & 0.5   & 0     & 2.6   & 8.8   & 2.7   & 7.4   & 7.1   & 2.8   & 10.7  & 1.4   & 7.9   & 1.6   & 6.1   & 6     & 1.8   & 9.8 \\
          &       &       & 0.5   & 4.6   & 10.6  & 4.7   & 7.3   & 7.2   & 4.6   & 7.8   & 5     & 10.2  & 5     & 5.8   & 6.8   & 4.6   & 8.9 \\
          &       & 0.8   & 0     & 8.1   & 16.5  & 7     & 12.8  & 15.2  & 8.4   & 20    & 4.7   & 10    & 4     & 7.2   & 9.7   & 5.4   & 13.5 \\
          &       &       & 0.5   & 13.4  & 17.8  & 12.9  & 13.2  & 17.2  & 14    & 17.7  & 11    & 11.6  & 12.9  & 9.6   & 11.3  & 11.7  & 13.8 \\
          & 0.001 & 0     & 0     & 13.3  & 21.4  & 10.2  & 13.8  & 9.5   & 14    & 11.2  & 41.6  & 38    & 39.7  & 35    & 35.4  & 42.2  & 37.5 \\
          &       &       & 0.5   & 17    & 22.1  & 14.6  & 13.3  & 11.5  & 14.3  & 12.5  & 46.2  & 39.4  & 44.5  & 40.4  & 41.4  & 45.6  & 41.6 \\
          &       & 0.5   & 0     & 5.3   & 11.4  & 4.4   & 8.9   & 8.8   & 4.5   & 13.2  & 8.5   & 22.1  & 10.4  & 17.4  & 14.9  & 11.7  & 20.3 \\
          &       &       & 0.5   & 5.3   & 13.8  & 5.9   & 9.8   & 8.1   & 6.5   & 14.8  & 18.1  & 24.7  & 18.2  & 20.1  & 22.8  & 20.3  & 27 \\
          &       & 0.8   & 0     & 8.3   & 17.5  & 7.7   & 14    & 14.9  & 8.1   & 19.3  & 8.1   & 16.7  & 7.8   & 12.8  & 13    & 5.8   & 16.8 \\
          &       &       & 0.5   & 14.4  & 19.2  & 15.4  & 14    & 16.9  & 15.4  & 20.6  & 15.6  & 14.8  & 16.6  & 16.4  & 15.3  & 15.7  & 18 \\
          & 0.01  & 0     & 0     & 38.9  & 37.3  & 34.6  & 31.8  & 34.3  & 37.3  & 32    & 78.3  & 72.7  & 74.9  & 67.7  & 79    & 80.3  & 73.3 \\
          &       &       & 0.5   & 40.2  & 46.2  & 37.7  & 35.3  & 40.2  & 44    & 39.9  & 83.8  & 77.2  & 79.4  & 75.3  & 79.8  & 84.5  & 79.7 \\
          &       & 0.5   & 0     & 10.9  & 23.8  & 9.9   & 17.7  & 15.5  & 11.5  & 25.1  & 29.3  & 40.1  & 29.4  & 36    & 38.4  & 32.1  & 46.4 \\
          &       &       & 0.5   & 18.1  & 25.8  & 12.8  & 16.5  & 19.2  & 17.4  & 25.2  & 41.5  & 37    & 37.9  & 39.1  & 44.8  & 40.1  & 45 \\
          &       & 0.8   & 0     & 12.1  & 23.7  & 10.9  & 16.7  & 20.1  & 11.2  & 22.1  & 18.1  & 30.1  & 18.2  & 22.7  & 26.9  & 17.9  & 31.9 \\
          &       &       & 0.5   & 19    & 24.9  & 16.4  & 19.3  & 21    & 21.1  & 25.9  & 35.3  & 30.9  & 33.2  & 28.4  & 32.9  & 32.9  & 36.3 \\
          & 0.1   & 0     & 0     & 65.1  & 68.1  & 62.7  & 61.8  & 68.4  & 63.4  & 64.1  & 96.6  & 94    & 95.1  & 91.8  & 96.2  & 96.4  & 95.2 \\
          &       &       & 0.5   & 68.5  & 71.7  & 67.8  & 64.2  & 72.4  & 71.4  & 66.8  & 96.4  & 93.2  & 95.8  & 93.8  & 96.5  & 96.7  & 96.2 \\
          &       & 0.5   & 0     & 25.7  & 36.5  & 22    & 28.9  & 31.4  & 27.7  & 41.2  & 47.1  & 43.6  & 48.1  & 43    & 55.3  & 49    & 53.7 \\
          &       &       & 0.5   & 33    & 36.1  & 25.4  & 28.2  & 36.2  & 33.1  & 42.2  & 49.3  & 43.6  & 49.5  & 41.8  & 57.3  & 51.6  & 55.1 \\
          &       & 0.8   & 0     & 26.4  & 30.8  & 25    & 26.5  & 29.2  & 25.6  & 36.9  & 41.3  & 41    & 40.1  & 41.2  & 48.4  & 42.8  & 50.5 \\
          &       &       & 0.5   & 32.3  & 33.8  & 28.9  & 27.6  & 32.8  & 33.2  & 39.7  & 52.2  & 44.6  & 49.1  & 39.7  & 52.7  & 48.8  & 52.7 \\
    \bottomrule
    \end{tabular}%
  \label{tab:sieve}%
			\end{scriptsize}
\end{table}%

\section{Empirical Applications}\label{sec:application}

\subsection{Environmental Kuznets curve}\label{subsec:EKC}

We first discuss an application of cointegrating polynomial regressions to the environmental Kuznets curve (EKC). It relates per capita GDP and, e.g., per capita $\mathrm{CO}_2$ emissions. The term EKC refers to the inverse U-shape relation of economic development and income inequality postulated by \cite{Kuznets1955}. It is motivated by the idea that both poor and rich countries emit little per capita, the first group because it does not yet resort to much heavily polluting activity such as individual car traffic, and the latter because it has access to more efficient technologies. The heavily (per capita) polluting middle-income group then gives rise to an inverse U-shape. \cite{Grossman1995} initiated a very active literature with contributions in several directions. See \cite{Stern2004,Stern2018} for more recent surveys.

We build on \cite{Wagner2015} and \cite{Stypka2017} who argued that using an ordinary \cite{Shin1994}-type linear cointegration test is inappropriate for cointegrating polynomial regressions (CPR). This is because the $k$-th power $x_t^k$ of an integrated regressor is not $I(1)$ anymore and thus violates the assumptions on the regressors of the \cite{Shin1994} test. Based on \cite{Wagner2016} the aforementioned authors applied a fully modified OLS approach for CPRs.
However, they did not consider variance breaks in their approach, which could lead to erroneous inference regarding the EKC hypothesis. We apply the bootstrap proposed in Section \ref{subsec:bootstrap} to address this possible issue in the following.

We study data of 19 industrialized countries from 1870 to 2014 (see Table \ref{tab:countries}; for New Zealand, data is available for 1878-2014). We use per capita GDP data of the Maddison database (\url{https://www.rug.nl/ggdc/historicaldevelopment/maddison/}). $\mathrm{CO}_2$ data is taken from the homepage of the Carbon Dioxide Information Analysis Center (\url{https://cdiac.ess-dive.lbl.gov/}) and is expressed as 1,000 tons per capita. We convert all time series to natural logarithms. Among others, \cite{Wagner2015} also examined sulfur dioxide data, but discussion and results are similar. For brevity, we only focus on (the arguably more relevant) $\mathrm{CO}_2$ emissions. Let $x_t$ denote log per capita GDP and $y_t$  log $\mathrm{CO}_2$ emissions per capita. We then study the model
\begin{equation*}
y_t=\delta_0+\delta_1 t+\theta_1x_t+\theta_2x_t^2+\theta_3x_t^3+u_t.
\end{equation*}

To assess whether variance breaks are present in the error term we follow \cite{Cavaliere2008} and define the \emph{empirical variance profile} as
\begin{equation}\label{eq:empvp}
\hat{\rho}(s):=\frac{\sum_{t=1}^{\left\lfloor Ts\right\rfloor}\hat{u}_t^2+(sT-\left\lfloor Ts\right\rfloor)\hat{u}_{\left\lfloor Ts\right\rfloor+1}^2}{\sum_{t=1}^{T}\hat{u}_t^2}
\end{equation}
for $s\in(0,1)$, with $\hat{\rho}(0):=0$ and $\hat{\rho}(1):=1$. In case of homoskedasticity, we should have $\hat{\rho}(s)\approx s$. Figure \ref{fig:vp1} plots the empirical variance profile for Australia, Austria, Belgium and Canada against $s$. Figures \ref{fig:vp2}--\ref{fig:vp5} for the remaining countries are given in the online appendix. We observe the presence of variance breaks for all countries (except maybe Denmark). For example, there is an early downward variance break for Canada. Thus, heteroskedasticity-robust tests are advisable.

\begin{figure}
\caption{Empirical variance profile \eqref{eq:empvp} for different countries. The dashed line is the reference line for homoskedasticity.}
\label{fig:vp1}
\begin{minipage}[t]{\textwidth}
	\centering
\include{varianceprofile1}
\end{minipage}
\end{figure}

Next, we run some univariate tests to characterize the series. In particular, we test for stationarity using a KPSS test (with the null of no unit root). Note that heteroskedasticity is an issue for the KPSS test as well, making critical values derived by \cite{KPSS1992} invalid. A possible remedy is to proceed as in \cite{Cavaliere2005}. We instead use our proposed bootstrap for the series $y_t$ and $x_t$ for residuals to test if they have no unit root and report the bootstrap $p$-values.

We perform two tests for cointegration, the bootstrap test using NLS residuals and the bootstrap test using DNLS residuals. We use a non-parametric autocorrelation-robust estimator for the variance with a Bartlett kernel and a spectral window of $\left\lfloor 4(T/100)^{0.25}\right\rfloor $ as suggested in \cite{KPSS1992}.

Table \ref{tab:countries} reports the test results for the different countries given in the first column. The second and third columns are for the bootstrap NLS and DNLS cointegration tests. Columns 5 and 6 give results for the KPSS test for $y_t$ and $x_t$. All test results are given by the corresponding $p$-values where very small $p$-values are abbreviated as $<.001$.

For the common level of significance of 5\% we draw the following conclusions: the KPSS test leads to a rejection of the null of no unit root of both $y_t$ and $x_t$ in all cases. This provides evidence that the regressor and the regressand are both $I(1)$. We also perform the non-robust \cite{Shin1994} test in column 4 of Table \ref{tab:countries} to test for a linear cointegrating relation. For most countries this hypothesis is rejected.

The two nonlinear cointegration tests reveal mixed results. The first observation is that both lead to acceptance of the null in the majority of the cases. Of course, bootstrap tests are dependent on simulation and the $p$-values are all close to the nominal size, so that decisions may hinge on simulation variability. To reduce the effects of randomness we increased the number of bootstrap runs to 2,000. The bootstrap tests yield conflicting test results in the case of Canada, Germany, Japan and Switzerland. Both tests reject for Australia, New Zealand, Portugal and the United States. In the other cases both tests accept the null, providing some support for the EKC hypothesis. \cite{Wagner2015} rejected the null for the majority of countries using fully modified OLS for cointegrating polynomial regressions. However, tests which are not robust to variance breaks can lead to size distortions.\footnote{The results are, in any case, not directly comparable since the Maddison database had a major update since then and also, since polynomials are sensitive to even small changes in the data.}

\begin{table}[tbp]
  \centering
  \caption{$p$-values for different tests. $p_{NLS}^b$ gives the $p$-value for the bootstrap NLS-based test and $p_{DNLS}^b$ for the bootstrap DNLS version, $p_{Shin}$ for the Shin test, $p_{KPSS,y}$ for the KPSS test for the $\mathrm{CO}_2$ emissions, $p_{KPSS,x}$ for the KPSS test for the GDP.}
    \begin{tabular}{lrrrrrr}
    \toprule
    Country & $p_{NLS}^b$ & $p_{DNLS}^b$ & $p_{Shin}$ & $p_{KPSS,y}$ & $p_{KPSS,x}$   \\
    \midrule
    Australia 			& .035	& .032	&	.039	& $<.001$		& $<.001$			 \\
    Austria 				& .390	& .366	&	.165	& $<.001$		& $<.001$			 \\
    Belgium 				& .560 	& .474	&	.039	& $<.001$		& $<.001$				 \\
    Canada 					& .053 	& .043	&$<.001$& $<.001$		& $<.001$				  \\
    Denmark 				& .097 	& .142	&	.011	& $<.001$		& $<.001$				  \\
    Finland 				& .251 	& .187	&	.008	& $<.001$		& $<.001$			  \\
    France 					& .186 	& .163	&	.003	& $<.001$		& $<.001$				  \\
    Germany 				& .027 	& .068	&	.006	& $<.001$		& $<.001$			  \\
    Italy						& .134 	& .143	&	.012	& $<.001$		& $<.001$			  \\
    Japan 					& .061 	& .019	&	.013	& $<.001$		& $<.001$				  \\
    Netherlands 		& .329 	& .276	&	.015	& $<.001$		& $<.001$				  \\
    New Zealand 		& .024 	& .027	&	.089	& $<.001$		& $<.001$				  \\
    Norway 					& .090	& .103	&	.052	& $<.001$		& $<.001$				  \\
    Portugal 				& .016 	& .026	&	.041	& $<.001$		& $<.001$				  \\
    Spain 					& .191 	& .113	&	.096 	& $<.001$		& $<.001$				  \\
    Sweden 					& .538 	& .432	&	.003	& $<.001$		& $<.001$				  \\
    Switzerland 		& .049 	& .053	&	.030	& $<.001$		& $<.001$				 \\
    United Kingdom 	& .174 	& .111	&	.040	& $<.001$		& $<.001$				  \\
    United States 	& .042 	& .037	&	.001	& $<.001$		& $<.001$			  \\
    \bottomrule
    \end{tabular}%
  \label{tab:countries}%
\end{table}%

\subsection{US money demand equation}\label{subsec:smoothapplication}
We next revisit the application in \cite{ChoiSaikkonen2010} who tested for a nonlinear cointegrating relation between money and the interest rate. In particular, allowing for a nonlinear adjustment process appears useful here: when the interest rate is high, the opportunity cost
of holding money increases, but it appears conceivable that the public only becomes sensitive when deviations are relevant. We contribute to this discussion by using our heteroskedasticity-robust test. We use the quarterly data from 1989 to 2016 from the International Financial Statistics. The data contains the four series M1 for money, GDP, the GDP deflator for the price level, and the 90-day Treasury bill rate for a short-term interest rate. M1 and GDP are seasonally adjusted. As in \cite{ChoiSaikkonen2010} we test for a cointegrating relation given by a smooth transition function. In order to do so, we transform the data to obtain the variables
$y_t=\log(\mathrm{M1}_t)-\log(\mathrm{GDP}\ \mathrm{deflator}_t)$, $x_{1t}=\log(\mathrm{GDP}_t)-\log(\mathrm{GDP}\ \mathrm{deflator}_t)$, and $x_{2t}=\log(\mathrm{Tbill}\ \mathrm{rate}_t)$. We use the model
\begin{equation*}
y_t=\delta_0+\theta_1x_{1t}+\theta_2x_{2t}+\theta_3\frac{1}{1+\exp(-\theta_4(x_t-\theta_3))}+u_t.
\end{equation*}

As in Subsection \ref{subsec:EKC}, we discuss the empirical variance profile \eqref{eq:empvp} to check for variance breaks. Figure \ref{fig:vps} reveals the likely presence of an upward variance break. Thus, the usage of heteroskedasticity-robust tests is advisable.
\begin{figure}
\caption{Empirical variance profile \eqref{eq:empvp} for the US money demand equation. The dashed line is the reference line for homoskedasticity.}
\label{fig:vps}
\begin{minipage}[t]{\textwidth}
	\centering
\include{varianceprofilesmooth}
\end{minipage}
\end{figure}

Table \ref{tab:money} presents $p$-values for the bootstrap-NLS and the bootstrap-DNLS tests. Again we have simulated $B=2,000$ bootstrap samples. Neither test rejects the null of cointegration for the 5\% level, which supports the findings of \cite{ChoiSaikkonen2010}.

\begin{table}[tbp]
  \centering
  \caption{$p$-values for different bootstrap tests with $B=2,000$ replications. $p_{NLS}^b$ gives the $p$-value for the bootstrap NLS-based test and $p_{DNLS1}^b$ for the bootstrap DNLS version with $K=1$, $p_{DNLS2}^b$ for the bootstrap DNLS version with $K=2$, $p_{DNLS3}^b$ for the bootstrap DNLS version with $K=3$.}
    \begin{tabular}{lrrrrr}
    \toprule
    $p_{NLS}^b$ 			& .226			 \\
    $p_{DNLS1}^b$			& .186			 \\
    $p_{DNLS2}^b$				& .206			 \\
    $p_{DNLS3}^b$					& .227				  \\
    \bottomrule
    \end{tabular}%
  \label{tab:money}%
\end{table}%

\section{Concluding remarks}\label{sec:conclusion}

This paper provides a test of the null of cointegration addressing a variety of features regularly arising in empirical applications. In particular, it simultaneously tackles nonlinearity, endogeneity, serial correlation and unconditional heteroskedasticity, thus providing a fair degree of generality. For example, the environmental Kuznets curve is a leading empirical model involving a nonlinear relationship. Next, as regressors of cointegration relationships can typically not be characterized as pure random walks nor equilibrium errors as pure white noise and moreover, as such series typically are correlated, allowing for serial correlation and endogeneity is, likewise, empirically relevant. Finally, phenomena such as the variance breaks arising from the Great Moderation highlight the need for inferential procedures robust to unconditional heteroskedasticity. If not properly accounted for, all these empirical features affect limiting distributions of test statistics and hence may render inference invalid.

We build on the KPSS-type test statistic for the null of cointegration of \citet{Shin1994}. One key building block for our approach is the work of \citet{ChoiSaikkonen2010}, based on which we tackle nonlinearity via suitable nonlinear least squares approaches, as well as endogeneity and serial correlation with dynamic OLS, also known as leads-and-lags regression. In turn, we address unconditional heteroskedasticity via a bootstrap approach that suitable handles the fact that different patterns of heteroskedasticity imply different null distributions, making conventional tabulation of critical values impractical. Concretely, we draw on the fixed-regressor wild bootstrap of \citet{CavaliereTaylor2006} and establish its validity in the present, more general framework.

Finally, Monte-Carlo simulations and empirical applications illustrate scenarios in which our proposal may be useful to practitioners. While the performance is in general satisfactory, we find, as expected and in line with related work in this literature, that, e.g., strong degrees of serial correlation and endogeneity have a detrimental impact on the performance of the proposed test.

\section*{Acknowledgements}

Financial support of the German Research Foundation (Deutsche Forschungsgemeinschaft, DFG) via the Collaborative Research Centers  SFB 823, project A4 and TRR 391, project C01 is gratefully acknowledged.

The authors thank four anonymous referees as well as Matei Demetrescu for comments that helped substantially improve the paper, as well as Janine Langerbein for excellent research assistance. There are no competing interests to declare.

\bibliographystyle{apalike}

\newpage
\appendix

\section{Proofs}\label{app:proofs}

\begin{proof}[Proof of Lemma \ref{lem:invprinc}]
The proof is an extension of \cite{CavaliereTaylor2006}. First recall that $\zeta_t=\Sigma_t^{1/2}\zeta_t^*$,  where $\{\zeta_t^*\}$ is stationary with zero-mean and unit variance which has no variance breaks. We recall the long-run covariance matrix
$\Gamma=\sum_{j=-\infty}^{\infty}E\left(\zeta_t^*(\zeta_{t-j}^*) ^{\prime}\right)$,
for any $t$. Thus the multivariate invariance principle \cite[]{Hansen1992} holds, i.e.,
\begin{equation*}
T^{-1/2}\sum_{t=1}^{\left\lfloor Ts\right\rfloor}\zeta_t^*\stackrel{w}{\rightarrow}\Gamma^{1/2}B(s),
\end{equation*}
where $B$ is a standard Brownian motion.

Next, recall from Assumption \ref{ass:Omega} that $\Sigma_T(s)=\Sigma(s)$. Since $T^{-1/2}\sum_{t=1}^{\left\lfloor Ts\right\rfloor}\zeta_t^*$ is independent of $\Sigma_T(s)$ joint convergence $\left(\Sigma_T(s),T^{-1/2}\sum_{t=1}^{\left\lfloor Ts\right\rfloor}\zeta_t^*\right)\stackrel{w}{\rightarrow}\left(\Sigma(s),\Gamma^{1/2}B(s)\right)$ follows, see \cite{Bilingsley1968}. Finally, by \cite{Hansen1992}, we obtain that for $0\le s\le1$
\begin{equation*}
T^{-1/2}\sum_{t=1}^{\left\lfloor Ts\right\rfloor}\zeta_t \stackrel{w}{\rightarrow}\int_0^s\Sigma^{1/2}(s)\Gamma^{1/2}\mathrm{d}B(s)=\int_0^s\Omega^{1/2}(s)\mathrm{d}B(s).
\end{equation*}
\end{proof}

\begin{proof}[Proof of Proposition \ref{prop:etaasymp}]
Consider $T^{-1/2}\sum_{t=1}^{\left\lfloor Ts\right\rfloor}\hat{u}_t$. Since $\hat{u}_t=u_t-(h(t_T,x_{tT},\hat{\vartheta}_T)-h(t_T,x_{tT},\vartheta_0))$, a second-order Taylor expansion of $h(t_T,x_{tT},\hat{\vartheta}_T)$ around $\vartheta_0$ gives
\begin{align}
T^{-1/2}\sum_{t=1}^{\left\lfloor Ts\right\rfloor}\hat{u}_t&= T^{-1/2}\sum_{t=1}^{\left\lfloor Ts\right\rfloor}u_t-T^{-1/2}\sum_{t=1}^{\left\lfloor Ts\right\rfloor}K(t_T,x_{tT},\vartheta_0)^{\prime}(\hat{\vartheta}_T-\vartheta_0)\label{eq:proof1}\\
&\quad+T^{1/2}(\hat{\vartheta}_T-\vartheta_0)^{\prime}\left(T^{-1}\sum_{t=1}^{\left\lfloor Ts\right\rfloor}\frac{\partial^2h(t_T,x_{tT},\tilde{\vartheta})}{\partial\vartheta\partial\vartheta^{\prime}}\right)(\hat{\vartheta}_T-\vartheta_0),\notag
\end{align}
where $||\tilde{\vartheta}-\vartheta_0||\le||\hat{\vartheta}_T-\vartheta_0||$. The generalized invariance principle (Lemma \ref{lem:invprinc}) implies that $\max_{1\le t\le T}||x_{tT}||=O_p(1)$ and since the function $\partial^2 h(\cdot,\cdot,\cdot)/\partial \vartheta\partial\vartheta^{\prime}$ is bounded on compact subsets of its domain, it follows by Lemma 1(i) of \cite{SaikkonenChoi2004} that the matrix in the middle of the third term on the RHS of \eqref{eq:proof1} is of order $O_p(1)$ uniformly in $0\le s\le1$. Combining this with Proposition \ref{prop:NLLasymp}, we obtain
\begin{equation}
\label{eq:proof1r}
T^{-1/2}\sum_{t=1}^{\left\lfloor Ts\right\rfloor}\hat{u}_t=T^{-1/2}\sum_{t=1}^{\left\lfloor Ts\right\rfloor}u_t-T^{-1/2}\sum_{t=1}^{\left\lfloor Ts\right\rfloor}K(t_T,x_{tT},\vartheta_0)^{\prime}(\hat{\vartheta}_T-\vartheta_0)+o_p(1),
\end{equation}
uniformly in $0\le s\le1$.

For the first term in \eqref{eq:proof1r} Lemma \ref{lem:invprinc} gives that, under $H_0$,
\begin{equation*}
T^{-1/2}\sum_{t=1}^{\left\lfloor Ts\right\rfloor}u_t=T^{-1/2}\sum_{t=1}^{\left\lfloor Ts\right\rfloor}\zeta_{u,t}\stackrel{w}{\rightarrow}B_{u,\Omega}(s).
\end{equation*}

For the second term in \eqref{eq:proof1r}, recall that $T^{1/2}(\hat{\vartheta}_T-\vartheta_0)\stackrel{w}{\rightarrow}\psi\left(B_{x,\Omega}^0,\vartheta_0,\kappa\right)$ (Proposition \ref{prop:NLLasymp}). By Lemma \ref{lem:invprinc},

\begin{equation*}
x_{tT}=(T_0/T)^{1/2}x_t=(T_0/T)^{1/2}\sum_{j=1}^{\left\lfloor Ts\right\rfloor}\zeta_{x,j}\stackrel{w}{\rightarrow}T_0^{1/2}B_{x,\Omega}(s)=:B_{x,\Omega}^0(s).
\end{equation*}
This implies that
\begin{equation*}
T^{-1}\sum_{t=1}^{\left\lfloor Ts\right\rfloor}x_{tT}\stackrel{w}{\rightarrow}\int_0^sB_{x,\Omega}^0(r)\mathrm{d}r
\end{equation*}
A standard result \cite[see, e.g.,][Proposition 17.1]{Hamilton1994} yields
\begin{equation}\label{eq:detconv}
T^{-1}\sum_{t=1}^{\left\lfloor Ts\right\rfloor}t_T^j\rightarrow T_0^j\frac{s^{1+j}}{1+j}=\int_0^s(T_0r)^j\mathrm{d}r
\end{equation}
for $0\le j\le q$.
It follows by the continuous mapping theorem that
\begin{equation}\label{eq:cmtKxtT}
T^{-1}\sum_{t=1}^{\left\lfloor Ts\right\rfloor}K(t_T,x_{tT},\vartheta_0)\stackrel{w}{\rightarrow}\int_0^sK(T_0r,B_{x,\Omega}^0(r),\vartheta_0)\mathrm{d}r=:F(s,B_{x,\Omega}^0,\vartheta_0).
\end{equation}

We conclude that
\begin{equation*}
T^{-1/2}\sum_{t=1}^{\left\lfloor Ts\right\rfloor}\hat{u}_t\stackrel{w}{\rightarrow}B_{u,\Omega}(s)-F(s,B_{x,\Omega}^0,\vartheta_0)^{\prime}\psi\left(B_{x,\Omega}^0,\vartheta_0,\kappa\right),
\end{equation*}
since all weak convergences hold jointly. Another application of the continuous mapping theorem yields
\begin{equation*}
T^{-2}\sum_{t=1}^T\left(\sum_{j=1}^t\hat{u}_j\right)^2\stackrel{w}{\rightarrow}\int_0^1\left(B_{u,\Omega}(s)-F(s,B_{x,\Omega}^0,\vartheta_0)^{\prime}\psi(B_{x,\Omega}^0,\theta_0,\kappa)\right)^2\mathrm{d}s.
\end{equation*}

Under the conditions of \cite{Andrews1991}, $\hat{\omega}_u^2$ is a consistent estimator of $\bar{\omega}_u^2$, as long as $T/l\to\infty$ for $T\to\infty$, see also \cite{Cavaliere2005}.

Finally, \eqref{eq:etaasymp} follows by the continuous mapping theorem.
\end{proof}

\begin{proof}[Proof of Theorem \ref{thm:etaLLasymp}]
Write \eqref{eq:LLresiduals} as
\begin{equation*}
\hat{e}_{Kt}=e_t-\left(h(t_T,x_{tT},\hat\vartheta_T^{(1)})-h(t_T,x_{tT},\vartheta_0)\right)+\sum_{|j|>K}\pi_j^{\prime}\zeta_{x,t-j}-V_t^{\prime}(\hat\pi_T^{(1)}-\pi_0).
\end{equation*}
As in \cite{ChoiSaikkonen2010} it is sufficient to show that the last two terms are asymptotically negligible. For the last,
\begin{equation}
\label{eq:asympnegligible1}
\max_{K+2\le n\le T-K}\left|N^{-1/2}\sum_{t=K+2}^n\left(\sum_{|j|>K}\pi_j^{\prime}\right)\right|=o_p(1)
\end{equation}
and
\begin{equation}\label{eq:asympnegligible2}
\max_{K+2\le n\le T-K}\left|N^{-1/2}\sum_{t=K+2}^nV_t^{\prime}(\hat\pi_T^{(1)}-\pi_0)\right|=o_p(1).
\end{equation}
\eqref{eq:asympnegligible1} follows by the same steps as in \cite{ChoiSaikkonen2010} except that we generalize their absolute summability of the autocovariance function for stationary errors to the summability $\sum_{l=-\infty}^{\infty}||E(\zeta_{x,t}\zeta_{x,t+l}^{\prime})||<\infty$, which is satisfied in view of \eqref{eq:summability}. \eqref{eq:asympnegligible2} follows analogously to \cite{ChoiSaikkonen2010}.

We obtain
\begin{align*}
N^{-1/2}\sum_{t=K+2}^{\left\lfloor (T-K)s\right\rfloor}\hat{e}_{Kt}=N^{-1/2}\sum_{t=K+2}^{\left\lfloor (T-K)s\right\rfloor}e_{t}-N^{-1/2}\sum_{t=K+2}^{\left\lfloor (T-K)s\right\rfloor}\left(h(t_T,x_{tT},\hat\vartheta_T^{(1)})-h(t_T,x_{tT},\vartheta_0)\right)+o_p(1),
\end{align*}
uniformly in $0\le s\le1$. As in the proof of Proposition \ref{prop:etaasymp}, we perform a second-order Taylor series expansion around $\vartheta_0$:
\begin{align*}
N^{-1/2}\sum_{t=K+2}^{\left\lfloor (T-K)s\right\rfloor}\hat{e}_{Kt}&= N^{-1/2}\sum_{t=K+2}^{\left\lfloor (T-K)s\right\rfloor}e_t-N^{-1/2}\sum_{t=K+2}^{\left\lfloor (T-K)s\right\rfloor}K(t_T,x_{tT},\vartheta_0)^{\prime}(\hat{\vartheta}_T^{(1)}-\vartheta_0)\\
&\quad+N^{1/2}(\hat{\vartheta}_T^{(1)}-\vartheta_0)^{\prime}\left(N^{-1}\sum_{t=K+2}^{\left\lfloor (T-K)s\right\rfloor}\frac{\partial^2h(t_T,x_{tT},\tilde{\vartheta})}{\partial\vartheta\partial\vartheta^{\prime}}\right)(\hat{\vartheta}_T^{(1)}-\vartheta_0),
\end{align*}
where $||\tilde{\vartheta}-\vartheta_0||\le ||\vartheta_T^{(1)}-\vartheta_0||$. As in the proof of Proposition \ref{prop:etaasymp},
\begin{equation*}
N^{-1/2}\sum_{t=K+2}^{\left\lfloor (T-K)s\right\rfloor}\hat{e}_{Kt}= N^{-1/2}\sum_{t=K+2}^{\left\lfloor (T-K)s\right\rfloor}e_t-N^{-1/2}\sum_{t=K+2}^{\left\lfloor (T-K)s\right\rfloor}K(t_T,x_{tT},\vartheta_0)^{\prime}(\hat{\vartheta}_T^{(1)}-\vartheta_0)+o_p(1),
\end{equation*}
uniformly in $0\le s\le1$. To conclude the statement, the first term converges to $B_{e,\omega}$. This holds by the invariance principle although $e_t$ is not necessarily strongly mixing. The reason for this is that, following \cite{SaikkonenChoi2004}, we can write $e_t=a^{\prime}(L)\left(\zeta_{u,t},\zeta_{x,t}^{\prime}\right)^{\prime}$, where $a^{\prime}(L)=\sum_{j=-\infty}^{\infty}a_j^{\prime}L^j=\left(1, -\pi(L)^{\prime}\right)$ and $\pi(L)=\sum_{j=-\infty}^{\infty}\pi_jL^j$. The summability condition \eqref{eq:pisumability} allows us to apply Theorem 4.2 of \cite{Saikkonen1993} and we obtain weak convergence to $B_{e,\omega}$. The remainder converges by the continuous mapping theorem and Proposition \ref{prop:LLasymp}.
\end{proof}

To prove Theorem \ref{thm:aympalternative} we need the following lemma. For completeness, we also prove the analogous statement for the NLS estimator.
\begin{lemma}\label{lem:aympalternative}
Under the alternative $H_1:\rho_{\mu}^2>0$,
\begin{equation*}
|\hat{\vartheta}_T-\vartheta_0|=O_p(T^{1/2})\ \ \ \mathrm{and}\ \ \ |\hat{\vartheta}_T^{(1)}-\vartheta_0|=O_p(T^{1/2}).
\end{equation*}
\end{lemma}

\begin{proof}
First, we discuss the NLS estimator. Under the alternative $H_1:\rho_{\mu}^2>0$
\begin{equation*}
T^{-1/2}u_{\left\lfloor Ts\right\rfloor}=T^{-1/2}\zeta_{u,\left\lfloor Ts\right\rfloor}+T^{-1/2}\rho_{\mu}\sum_{t=1}^{\left\lfloor Ts\right\rfloor}\zeta_{\mu,t}\stackrel{w}{\rightarrow}\rho_{\mu}B_{\mu,\Omega}(s).
\end{equation*}
This implies that $T^{-3/2}\sum_{t=1}^{\left\lfloor Ts\right\rfloor}u_t\stackrel{w}{\rightarrow}\rho_{\mu}\int_0^{s}B_{\mu,\Omega}(r)\mathrm{d}r$ and hence $T^{-3/2}\sum_{t=1}^{\left\lfloor Ts\right\rfloor}u_t=O_p(1)$. Moreover, the joint convergence, Lemma \ref{lem:invprinc} and the continuous mapping theorem yield
\begin{align}
T^{-3/2}\sum_{t=1}^{\left\lfloor Ts\right\rfloor} K(t_T,x_{tT},\vartheta_0)u_t&=T^{-3/2}\sum_{t=1}^{\left\lfloor Ts\right\rfloor} K(t_T,x_{tT},\vartheta_0)\left(\sum_{j=1}^t\rho_{\mu}\zeta_{\mu,j}+\zeta_{u,t}\right)\label{eq:NLLasympH1bla}\\
&=T^{-3/2}\sum_{t=1}^{\left\lfloor Ts\right\rfloor} K(t_T,x_{tT},\vartheta_0)\sum_{j=1}^t\rho_{\mu}\zeta_{\mu,j}+o_p(1)\notag\\
&\stackrel{w}{\rightarrow} \int_0^1K\left(T_0s,B_{x,\Omega}^0(s),\vartheta_0\right)\rho_{\mu}\mathrm{d}B_{\mu,\Omega}(s).\notag
\end{align}
Note that we have used the assumption that $\zeta_{x,0}$ and $\zeta_{\mu,t}$ are uncorrelated.

We now discuss a similar statement to \eqref{eq:NLLasymp} under $H_1$. More precisely,
\begin{align*}
T^{-1/2}\left(\hat{\vartheta}_T-\vartheta_0\right)\stackrel{w}{\rightarrow}&\left(\int_0^1K\left(T_0s,B_{x,\Omega}^0(s),\vartheta_0\right)K\left(T_0s,B_{x,\Omega}^0(s),\vartheta_0\right)^{\prime}\mathrm{d}s\right)^{-1}\\&\times\int_0^1K\left(T_0s,B_{x,\Omega}^0(s),\vartheta_0\right)\rho_{\mu}\mathrm{d}B_{\mu,\Omega}(s).
\end{align*}
To obtain this we adopt the proof of Theorem 2 in \cite{SaikkonenChoi2004} and replace $T^{-1/2}\sum_{t=1}^{\left\lfloor Ts\right\rfloor} K(t_T,x_{tT},\vartheta_0)u_t$ (in our notation) by $T^{-3/2}\sum_{t=1}^{\left\lfloor Ts\right\rfloor} K(t_T,x_{tT},\vartheta_0)u_t$ in (B.3) of \cite{SaikkonenChoi2004}.

Second, for the DNLS estimator,
\begin{align*}
N^{-1/2}e_{K\left\lfloor (T-K)s\right\rfloor}&=N^{-1/2}e_{\left\lfloor (T-K)s\right\rfloor}+N^{-1/2}\sum_{|j|>K}\pi_j^{\prime}\zeta_{x,\left\lfloor (T-K)s\right\rfloor-j}\\
&=N^{-1/2}u_{\left\lfloor (T-K)s\right\rfloor}-N^{-1/2}\sum_{j=-K}^K\pi_j^{\prime}\zeta_{x,\left\lfloor (T-K)s\right\rfloor-j}\\
&=N^{-1/2}\rho_{\mu}\sum_{t=1}^{\left\lfloor (T-K)s\right\rfloor}\zeta_{\mu,t}+o_p(1).
\end{align*}
The remainder follows analogously to \eqref{eq:NLLasympH1bla} above.
\end{proof}

\begin{proof}[Proof of Theorem \ref{thm:aympalternative}]

Like in the proofs of Proposition \ref{prop:etaasymp} and Theorem \ref{thm:etaLLasymp} we use a Taylor expansion to obtain
\begin{equation}\label{eq:Taylorexpalternative}
T^{-3/2}\sum_{t=1}^{\left\lfloor Ts\right\rfloor}\hat{u}_t= T^{-3/2}\sum_{t=1}^{\left\lfloor Ts\right\rfloor}u_t-T^{-3/2}\sum_{t=1}^{\left\lfloor Ts\right\rfloor}K(t_T,x_{tT},\vartheta_0)^{\prime}(\hat{\vartheta}_T-\vartheta_0)+o_p(1),
\end{equation}
and
\begin{equation}\label{eq:TaylorexpalternativeLL}
N^{-3/2}\sum_{t=K+2}^{\left\lfloor (T-K)s\right\rfloor}\hat{e}_{Kt}= N^{-3/2}\sum_{t=K+2}^{\left\lfloor (T-K)s\right\rfloor}e_t-N^{-3/2}\sum_{t=K+2}^{\left\lfloor (T-K)s\right\rfloor}K(t_T,x_{tT},\vartheta_0)^{\prime}(\hat{\vartheta}_T-\vartheta_0)+o_p(1).
\end{equation}

Next, we use Lemma \ref{lem:aympalternative} to get $|\hat{\vartheta}_T-\vartheta_0|=O_p(T^{1/2})$, $|\hat{\vartheta}_T^{(1)}-\vartheta_0|=O_p(T^{1/2})$ and \eqref{eq:cmtKxtT} to get $\sum_{t=1}^{\left\lfloor Ts\right\rfloor}K(t_T,x_{tT},\vartheta_0)=O_p(T)$, which implies that the second terms on the RHS of \eqref{eq:Taylorexpalternative} and \eqref{eq:TaylorexpalternativeLL} resp., are $O_p(1)$.
Therefore, $\sum_{t=1}^{\left\lfloor Ts\right\rfloor}\hat{u}_t=O_p(T^{3/2})$ and $\sum_{t=K+2}^{\left\lfloor (T-K)s\right\rfloor}\hat{e}_{Kt}=O_p(T^{3/2})$, which leads to
\begin{equation*}
T^{-2}\sum_{t=1}^T\left(\sum_{j=1}^t\hat{u}_j\right)^2=O_p(T^2)
\end{equation*}
and
\begin{equation*}
N^{-2}\sum_{t=K+2}^{T-K}\left(\sum_{j=K+2}^t\hat{e}_{Kj}\right)^2=O_p(T^2).
\end{equation*}

Moreover, \cite{KPSS1992} showed that the long-run variance estimator satisfies $\hat{\omega}_u^2=O(lT)$, which carries over to $\hat{\omega}_e^2$. This implies $\hat{\eta}_{NLS}=O_p(T/l)$ and $\hat{\eta}_{DNLS}=O_p(T/l)$. As long as $T/l\to\infty$ for $T\to\infty$ the tests are consistent.
\end{proof}

\begin{proof}[Proof of Theorem \ref{thm:bootstrap}]
\begin{enumerate}[label=(\roman*)]
\item Similarly to the proof of Theorem 3 in \cite{CavaliereTaylor2006} consider the process $M_T^b$ s.th.
\begin{equation*}
M_T^b(s):=N^{-1/2}\sum_{t=K+2}^{\left\lfloor (T-K)s\right\rfloor}u_t^b=N^{-1/2}\sum_{t=K+2}^{\left\lfloor (T-K)s\right\rfloor}\hat{e}_{Kt}z_t.
\end{equation*}
Conditionally on $\{\hat{e}_{Kt},x_{tT}\}_{t=1}^T$, this is an exact Gaussian process with kernel
\begin{equation*}
\Lambda_T^M(s,s^{\prime})=N^{-1}\sum_{t=K+2}^{\left\lfloor (T-K)(s\wedge s^{\prime})\right\rfloor}\hat{e}_{Kt}^2,
\end{equation*}
where $s\wedge s^{\prime}$ denotes the minimum of $s$ and $s^{\prime}$.

Under the null, $Var(e_t)=\sigma_{e,t}^2$ and $\sigma_e^2(s)=\sigma_{e,\left\lfloor Ts\right\rfloor}^2$, the variance profile of $e_t$.
As in the proof of Lemma A.5 in \cite{Cavaliere2010} we see that
\begin{equation}\label{eq:ssq}
T^{-1}\sum_{t=1}^{\left\lfloor T(s\wedge s^{\prime})\right\rfloor}\hat{e}_{Kt}^2=T^{-1}\sum_{t=1}^{\left\lfloor T(s\wedge s^{\prime})\right\rfloor}e_t^2+o_p(1)\stackrel{p}{\rightarrow}\int_0^{s\wedge s^{\prime}}\sigma_e^2(r)\mathrm{d}r,
\end{equation}
pointwise, where the first equality follows by the proof of Theorem 2 in \cite{McCabe1997}. Since $T^{-1}\sum_{t=1}^{\left\lfloor Ts\right\rfloor}\hat{e}_{Kt}^2$ is monotonically increasing in $s$ and the limit function is continuous in $s$ the convergence in probability is also uniform. The RHS is the kernel of the Gaussian process $B_{e,\sigma}(s):=\int_{0}^{s}\sigma_e(r)\mathrm{d}B_e(r)$, where $B_e$ is a standard Brownian motion as above. This implies that $M_T^b(s)\stackrel{w}{\rightarrow}_p B_{e,\sigma}(s)$, as in \cite{Hansen1996}.

Analogously, applying the same mappings as in the proofs of Proposition \ref{prop:etaasymp} and Theorem \ref{thm:etaLLasymp},
\begin{equation*}
N^{-2}\sum_{t=K+2}^{T-K}\left(\sum_{j=K+2}^t\hat{e}_{Kj}^b\right)^2\stackrel{w}{\rightarrow}_p\int_0^1\left(B_{e,\sigma}(s)-F(s,B_{x,\Omega}^0,\vartheta_0)^{\prime}\chi_{\sigma}(B_{x,\Omega}^0,\vartheta_0)\right)^2\mathrm{d}s.
\end{equation*}

We now derive the large sample behavior of $(\hat{\omega}_{e}^b)^2$, which, thanks to the multiplication with wild bootstrap errors, reduces to that of \eqref{eq:ssq}.
\begin{eqnarray}
(\hat{\omega}_{e}^b)^2&=&N^{-1}\sum_{t=K+2}^{T-K}(\hat{e}_{Kt}^b)^2+2N^{-1}\sum_{s=K+2}^l w(s,l)\sum_{t=s+1}^{T-K}\hat{e}_{Kt}^b\hat{e}_{K(t-s)}^b\label{eq:omegaub2}\\
&=&N^{-1}\sum_{t=K+2}^{T-K}(e_t^b)^2+2N^{-1}\sum_{s=K+2}^lw(s,l)\sum_{t=s+1}^{T-K}e_t^be_{t-s}^b+o_p(1)\notag\\
&\stackrel{p}{\rightarrow}&\int_0^{1}\sigma^2(r)\mathrm{d}r,\notag
\end{eqnarray}
because $E(z_tz_{t-s}|\{\hat{e}_{Kt},x_{tT}\}_{t=1}^T)$ equals 0 for all $s>0$ and equals 1 for $s=0$, and the same argument as above by \cite{McCabe1997}.

\item We again consider $M_T^b(s)$ and $\Lambda_T^M(s,s^{\prime})$ but now it suffices to look at the order of convergence. Recall that under the alternative $\sum_{t=K+2}^{\left\lfloor (T-K)s\right\rfloor}\hat{e}_{Kt}=O_p(T^{3/2})$ and $\sum_{t=K+2}^{\left\lfloor (T-K)s\right\rfloor}\hat{e}_{Kt}^2=O_p(T^{2})$. This implies that $\Lambda_T^M(s,s^{\prime})=O_p(T)$ and, like in part (i), $T^{-1/2} M_T^b(s)$ converges weakly in probability to a Gaussian process where the kernel is given by the weak limit of $T^{-1}\Lambda_T^M(s,s^{\prime})$.

By the continuous mapping theorem it follows that $\sum_{t=K+2}^{T-K}\hat{e}_{Kt}^b=O_p(T)$ and, hence, that
\begin{equation*}
\sum_{t=K+2}^{T-K}\left(\sum_{j=K+2}^t\hat{e}_{Kj}^b\right)^2=O_p(T^3).
\end{equation*}
Consider next the long-run variance estimator $(\hat{\omega}_{e}^b)^2$.
Going back to \eqref{eq:omegaub2}, we again note that $E(z_tz_{t-s}|\{\hat{e}_{Kt},x_{tT}\}_{t=1}^T)=0$ for all covariance terms $s\neq0$. This implies that the second term on the RHS of
\begin{equation*}
\frac{1}{N}(\hat{\omega}_{e}^b)^2=N^{-2}\sum_{t=K+2}^{T-K}(\hat{e}_{Kt}^b)^2+2N^{-2}\sum_{s=K+2}^lw(s,l)\sum_{t=s+1}^{T-K}\hat{e}_{Kt}^b\hat{e}_{K(t-s)}^b
\end{equation*}
converges to zero in probability. Thus, the analysis of the bootstrap long-run variance estimator reduces to the analysis of the bootstrap variance estimator. From \citet[p.~634]{CavaliereTaylor2006} it therefore follows that $(\hat{\omega}_{e}^b)^2/N$ converges weakly in probability; $N^{-2}\sum_{t=K+2}^{T-K}(\hat{e}_{Kt}^b)^2$ being, under the alternative, the scaled residual variance estimator of a spurious regression, see also \citet[eq.~(A.9)]{Phillips1986}. Hence $(\hat{\omega}_{e}^b)^2=O_p(T)$.
All in all, we get $\hat{\eta}_{DNLS}^b=O_p(1)$.
\end{enumerate}
\end{proof}

\begin{proof}[Proof of Corollary \ref{cor:bootstrap}]
\begin{enumerate}[label=(\roman*)]
\item The bootstrap test statistic $\hat{\eta}_{DNLS}^b$ samples from a distribution that has the same variance profile as the distribution of $\hat{\eta}_{DNLS}$ but with white noise serial correlation. In particular, Lemma 1 of \cite{Cavaliere2005} implies that $B_{e,\omega}(s)=\gamma_eB_{e,\sigma}(s)$, where $\gamma_e$ is the long-run variance in \eqref{eq:gammae}. Furthermore, $\omega_e(s)=\gamma_e\sigma_e(s)$. This also implies that $\chi_{\omega}(B_{x,\Omega}^0,\vartheta_0)$ in \eqref{eq:chiomegae} can be written as $\gamma_e\chi_{\sigma}(B_{x,\Omega}^0,\vartheta_0)$. Therefore, we can rearrange the RHS of \eqref{eq:etaLLasymp} to obtain
\begin{align*}
&\bar{\omega}_e^{-2}\int_0^1\left(B_{e,\omega}(s)-F(s,B_{x,\Omega}^0,\vartheta_0)^{\prime}\chi_{\omega}(B_{x,\Omega}^0,\vartheta_0)\right)^2\mathrm{d}s\\
=&\;\bar{\sigma}_e^{-2}\int_0^1\left(B_{e,\sigma}(s)-F(s,B_{x,\Omega}^0,\vartheta_0)^{\prime}\chi_{\sigma}(B_{x,\Omega}^0,\vartheta_0)\right)^2\mathrm{d}s.
\end{align*}
Therefore, the bootstrap distribution correctly replicates the asymptotic distribution. The probability integral transform (see the proof of Theorem 2 in \citet{Cavaliere+Taylor-BootUnitRootTest:08}) implies $p_T^b\stackrel{w}{\rightarrow}\mathcal{U}[0,1]$.
\item Since $\hat{\eta}_{DNLS}=O_p(T/l)$ under $H_1$ (Theorem \ref{thm:aympalternative}), Theorem \ref{thm:bootstrap} (ii) implies that $p_T^b\stackrel{p}{\rightarrow}0$, as long as $l/T\to0$ for $T\to \infty$.
\end{enumerate}
\end{proof}

\newpage
\begin{center}
\Large{Online Appendix for\\\medskip ``Testing for Nonlinear Cointegration under Heteroskedasticity''}\\\medskip\medskip
\large{Christoph Hanck and Till Massing}
\\\medskip
\large{University of Duisburg-Essen}
\end{center}

\section{Robustness against misspecification}\label{subsec:robustness}

We perform a robustness check where the specified model differs from the DGP. More precisely, we consider a cubic DGP
\begin{equation*}
y_t=x_t+2x_t^2+x_t^3+u_t,
\end{equation*}
whilst we misspecify the model for estimation and regress $y_t$ on $g(x_t,\theta)=\theta_1x_t+\theta_2x_t^2$. Table \ref{tab:robust} shows the empirical rejection frequencies for the usual parameter constellations. We observe that, unsurprisingly, a misspecified model leads to severe size distortions making the bootstrap test invalid. This seems intuitive in that misspecification generally is an issue for cointegration tests, and inferential procedures more generally. To address this issue we advise to perform RESET tests in which one includes higher powers and tests for cointegration in the extended model. We refer to Vogelsang and Wagner, 2014, An integrated modified {OLS} {RESET} test for cointegrating regressions (Working Paper), for a discussion using the integrated modified OLS estimator.

\begin{table}[bp]
    \begin{scriptsize}
		\caption{Empirical rejection frequencies for testing the null of cointegration in the using a \emph{quadratic} regression model with a \emph{cubic} DGP for various parameter constellations for the DNLS bootstrap test. All rejection rates are given as percentages. The nominal size is 5\%.}
  \hspace{-0.5cm}
    \begin{tabular}{crrrrrrrrrrrrrrrrr}
    	    \toprule
		& &       & \multicolumn{1}{r}{$T$} & \multicolumn{7}{c}{100}       & \multicolumn{7}{c}{300}    \\
		\cmidrule(l{.75em}){5-11}\cmidrule(l{.75em}){12-18}
     &     &       & \multicolumn{1}{r}{$\tau$} & 0     & \multicolumn{2}{c}{0.1}          & \multicolumn{2}{c}{0.5}   &        \multicolumn{2}{c}{0.9}    & 0     & \multicolumn{2}{c}{0.1}   &  \multicolumn{2}{c}{0.5}  &   \multicolumn{2}{c}{0.9}     \\
					\cmidrule(l{.75em}){6-7}\cmidrule(l{.75em}){8-9}\cmidrule(l{.75em}){10-11}\cmidrule(l{.75em}){13-14}\cmidrule(l{.75em}){15-16}\cmidrule(l{.75em}){17-18}
     &     &       & \multicolumn{1}{r}{$\omega$} &       & 1/16 & 16    & 1/16 & 16    & 1/16 & 16    &       & 1/16 & 16    & 1/16 & 16    & 1/16 & 16 \\
  &  \multicolumn{1}{r}{$\rho_{\mu}^2$} & \multicolumn{1}{r}{$\rho$} & \multicolumn{1}{r}{$\lambda$} &       &       &       &       &       &       &       &       &       &       &       &       &       &  \\ \midrule
       & 0     & 0     & 0     & 76.2  & 69.4  & 78.2  & 74.8  & 80.8  & 75.9  & 78    & 90.6  & 85.6  & 92    & 90.5  & 94    & 91.5  & 90.7 \\
          &       &       & 0.5   & 76.3  & 64.1  & 76.4  & 76.1  & 80.7  & 77.6  & 76.7  & 92.2  & 87.2  & 92.8  & 90.6  & 95.5  & 91    & 90.1 \\
          &       & 0.5   & 0     & 45    & 24.7  & 41.9  & 36.8  & 49.5  & 43.5  & 45.5  & 52.2  & 36.8  & 52.7  & 51.2  & 59.9  & 54.1  & 56.6 \\
          &       &       & 0.5   & 43.4  & 23.8  & 42.2  & 33.9  & 53.5  & 43    & 46.4  & 54.6  & 39.6  & 54.5  & 51    & 60.7  & 58    & 55.8 \\
          &       & 0.8   & 0     & 41.4  & 23.3  & 37.7  & 37.1  & 49    & 45.1  & 45.1  & 55.8  & 35.7  & 56.2  & 54.2  & 63.3  & 56.2  & 55.6 \\
          &       &       & 0.5   & 44.1  & 23.9  & 38.8  & 35.4  & 49.4  & 46.1  & 44.7  & 57.1  & 37.6  & 52.9  & 51.7  & 62.9  & 54.5  & 53.7 \\\vspace{0.15cm}
          & 0.001 & 0     & 0     & 78.1  & 67.6  & 76.7  & 75.8  & 81.8  & 78.2  & 76.4  & 91.3  & 86.4  & 89    & 90.3  & 93.8  & 91.4  & 90.1 \\
          &       &       & 0.5   & 78.3  & 68.3  & 76.3  & 75.1  & 81.8  & 77.5  & 76.1  & 91.6  & 86.3  & 90.8  & 92.1  & 94.8  & 92.4  & 88.9 \\
          &       & 0.5   & 0     & 47    & 26.6  & 42.8  & 38.5  & 53.3  & 41.7  & 45.1  & 56.9  & 37.1  & 52    & 50.3  & 61.6  & 54.4  & 53.7 \\
          &       &       & 0.5   & 45.2  & 27.7  & 40    & 37.1  & 48.7  & 44.8  & 41.7  & 55.1  & 36.5  & 54.1  & 51.4  & 64.4  & 56.4  & 55 \\
          &       & 0.8   & 0     & 43.5  & 24.3  & 42.4  & 35.2  & 50.9  & 41.9  & 41.4  & 56.5  & 37.7  & 53.5  & 49.9  & 64.3  & 53.5  & 54.3 \\
          &       &       & 0.5   & 43.5  & 26.6  & 38.9  & 34.8  & 52.3  & 46.8  & 41.9  & 53.7  & 36.9  & 55.2  & 53    & 63.2  & 56.7  & 53.1 \\
          & 0.01  & 0     & 0     & 76.6  & 71.1  & 76    & 75    & 81.1  & 78.9  & 76    & 92    & 84.8  & 92.3  & 90.3  & 94.1  & 92.5  & 89.1 \\
          &       &       & 0.5   & 77.5  & 68.8  & 73.4  & 76    & 84.2  & 74.8  & 75.9  & 91.1  & 86    & 90.4  & 90.2  & 93.8  & 91.3  & 89.5 \\
          &       & 0.5   & 0     & 42.7  & 28.3  & 41.3  & 34.8  & 50.2  & 43.6  & 43.8  & 57.8  & 37    & 56.6  & 50.2  & 62.1  & 56.1  & 52.2 \\
          &       &       & 0.5   & 43.4  & 27    & 40.1  & 36.2  & 56.6  & 42.2  & 41.4  & 54.3  & 36.8  & 53.5  & 51.2  & 63.7  & 56    & 54 \\
          &       & 0.8   & 0     & 41.7  & 22.4  & 39    & 35.3  & 52.9  & 43.4  & 42.4  & 56.6  & 34.5  & 55.4  & 53.2  & 63.2  & 56.5  & 54.6 \\
          &       &       & 0.5   & 41.9  & 24.6  & 41.9  & 39.3  & 50.9  & 40.8  & 45    & 55.1  & 36    & 50    & 51.1  & 63.5  & 54.2  & 55.5 \\
          & 0.1   & 0     & 0     & 79    & 69.5  & 78.1  & 76.3  & 82.1  & 78.5  & 76.4  & 90.3  & 87.8  & 92    & 92.1  & 95.3  & 92.4  & 89.3 \\
          &       &       & 0.5   & 77.6  & 69.8  & 74.3  & 73.5  & 81.8  & 76.8  & 74.9  & 92.1  & 86    & 91.5  & 90.4  & 93.8  & 93.2  & 90.5 \\
          &       & 0.5   & 0     & 45.2  & 29.2  & 41.4  & 35.6  & 53.4  & 44.3  & 45.1  & 55.4  & 37    & 52.4  & 52.5  & 64.7  & 55.1  & 53.8 \\
          &       &       & 0.5   & 44.1  & 27.9  & 41.1  & 33.7  & 53.1  & 42.1  & 42    & 57.4  & 34.7  & 53.8  & 50.4  & 63.2  & 55.6  & 54.5 \\
          &       & 0.8   & 0     & 44.7  & 26.3  & 40    & 36.6  & 51.6  & 40.3  & 42.1  & 53.8  & 35.7  & 54.5  & 49.2  & 62.3  & 55.1  & 55.9 \\
          &       &       & 0.5   & 44    & 28.2  & 42.6  & 32.2  & 52    & 43    & 44.5  & 54.6  & 36.8  & 52    & 50.5  & 62.7  & 56    & 54.2 \\
    \bottomrule
    \end{tabular}%
  \label{tab:robust}%
		\end{scriptsize}
\end{table}%

\newpage
\section{Additional Plots}\label{app:plots}

\begin{figure}
\caption{Empirical variance profile \eqref{eq:empvp} for different countries. The dashed line is the reference line for homoskedasticity.}
\label{fig:vp2}
\begin{minipage}[t]{\textwidth}
	\centering
\include{varianceprofile2}
\end{minipage}
\end{figure}

\begin{figure}
\caption{Empirical variance profile \eqref{eq:empvp} for different countries. The dashed line is the reference line for homoskedasticity.}
\label{fig:vp3}
\begin{minipage}[t]{\textwidth}
	\centering
\include{varianceprofile3}
\end{minipage}
\end{figure}

\begin{figure}
\caption{Empirical variance profile \eqref{eq:empvp} for different countries. The dashed line is the reference line for homoskedasticity.}
\label{fig:vp4}
\begin{minipage}[t]{\textwidth}
	\centering
\include{varianceprofile4}
\end{minipage}
\end{figure}

\begin{figure}
\caption{Empirical variance profile \eqref{eq:empvp} for different countries. The dashed line is the reference line for homoskedasticity.}
\label{fig:vp5}
\begin{minipage}[t]{\textwidth}
	\centering
\include{varianceprofile5}
\end{minipage}
\end{figure}

\end{document}